\documentclass[reprint,prd,nofootinbib,superscriptaddress]{revtex4}
\usepackage{amsfonts}
\usepackage{amsmath}
\usepackage{amssymb}
\usepackage[english]{babel}
\usepackage{graphicx}
\usepackage{epsfig}
\usepackage{bm}
\usepackage{verbatim}
\usepackage[utf8]{inputenc}
\usepackage{booktabs}
\usepackage{multirow}
\usepackage{xcolor}
\usepackage[colorlinks=true,urlcolor=red,citecolor=red]{hyperref}
\usepackage[font=small]{caption}
\usepackage{float}
\usepackage{blindtext}
\usepackage{booktabs}
\usepackage{subfigure}
\usepackage{tikz}
\usepackage{pgffor}

\usetikzlibrary{arrows,shapes}
\usetikzlibrary{trees}
\usetikzlibrary{matrix,arrows} 
\usetikzlibrary{positioning}
\usetikzlibrary{calc,through}
\usetikzlibrary{decorations.pathreplacing}  
\usetikzlibrary{decorations.pathmorphing}
\usetikzlibrary{decorations.markings}
\tikzset{
    vector/.style={decorate, decoration={snake}, draw},
provector/.style={decorate, decoration={snake,amplitude=2.5pt}, draw},
antivector/.style={decorate, decoration={snake,amplitude=-2.5pt}, draw},
    fermion/.style={draw=black, postaction={decorate},
        decoration={markings,mark=at position .55 with {\arrow[draw=black]{>}}}},
    fermionbar/.style={draw=black, postaction={decorate},
        decoration={markings,mark=at position .55 with {\arrow[draw=black]{<}}}},
    fermionnoarrow/.style={draw=black},
    gluon/.style={decorate, draw=black,
        decoration={coil,amplitude=4pt, segment length=5pt}},
    scalar/.style={dashed,draw=black, postaction={decorate},
        decoration={markings,mark=at position .55 with {\arrow[draw=black]{>}}}},
    scalarbar/.style={dashed,draw=black, postaction={decorate},
        decoration={markings,mark=at position .55 with {\arrow[draw=black]{<}}}},
    scalarnoarrow/.style={dashed,draw=black},
    electron/.style={draw=black, postaction={decorate},
        decoration={markings,mark=at position .55 with {\arrow[draw=black]{>}}}},
bigvector/.style={decorate, decoration={snake,amplitude=4pt}, draw},
}\usetikzlibrary{decorations.markings}
\tikzstyle{block} = [draw, rectangle, 
    minimum height=3em, minimum width=6em]

\newcommand{\be}{\begin{equation}}
\newcommand{\ee}{\end{equation}}
\newcommand{\bea}{\begin{eqnarray}}
\newcommand{\eea}{\end{eqnarray}}
\newcommand{\ben}{\begin{enumerate}}
\newcommand{\een}{\end{enumerate}}
\newcommand{\bde}{\begin{widetext}}
\newcommand{\ede}{\end{widetext}}
\newcommand{\nn}{\nonumber}

\newcommand{\al}{\alpha}
\newcommand{\la}{\lambda}

\newcommand{\bc}{\begin{center}}
\newcommand{\ec}{\end{center}}

\newcommand{\mathsym}[1]{}

\newcommand{\AC}[1]{{#1}}
\newcommand{\VKN}[1]{{#1}}
\newcommand{\Binh}[1]{{#1}}
\newcommand{\LH}[1]{{#1}}
\newcommand{\CB}[1]{{#1}}

\input{tcilatex}
\usepackage{graphicx}
\usepackage{bm}%para colocar ecuaciones en negrita, es especial las letras griegas.
\usepackage{float}
\usepackage{subfigure}
\usepackage{tikz}
\usetikzlibrary{arrows.meta}
\usetikzlibrary{shapes.geometric}

\usetikzlibrary{arrows,shapes}
\usetikzlibrary{trees}
\usetikzlibrary{matrix,arrows} 				% For commutative diagram
\usetikzlibrary{positioning}				% For "above of=" commands
\usetikzlibrary{calc,through}				% For coordinates
\usetikzlibrary{decorations.pathreplacing}  % For curly braces
\usepackage{pgffor}							% For repeating patterns

\usetikzlibrary{decorations.pathmorphing}	% For Feynman Diagrams
\usetikzlibrary{decorations.markings}

\tikzset{
    vector/.style={decorate, decoration={snake}, draw},
	provector/.style={decorate, decoration={snake,amplitude=2.5pt}, draw},
	antivector/.style={decorate, decoration={snake,amplitude=-2.5pt}, draw},
    fermion/.style={draw=black, postaction={decorate},
        decoration={markings,mark=at position .55 with {\arrow[draw=black]{>}}}},
    fermionbar/.style={draw=black, postaction={decorate},
        decoration={markings,mark=at position .55 with {\arrow[draw=black]{<}}}},
    fermionnoarrow/.style={draw=black},
    gluon/.style={decorate, draw=black,
        decoration={coil,amplitude=4pt, segment length=5pt}},
    scalar/.style={dashed,draw=black, postaction={decorate},
        decoration={markings,mark=at position .55 with {\arrow[draw=black]{>}}}},
    scalarbar/.style={dashed,draw=black, postaction={decorate},
        decoration={markings,mark=at position .55 with {\arrow[draw=black]{<}}}},
    scalarnoarrow/.style={dashed,draw=black},
    electron/.style={draw=black, postaction={decorate},
        decoration={markings,mark=at position .55 with {\arrow[draw=black]{>}}}},
	bigvector/.style={decorate, decoration={snake,amplitude=4pt}, draw},
}
\usetikzlibrary{decorations.markings}
\begin{document}

\title{Phenomenology of 3-3-1 models with radiative inverse seesaw mechanism}
\author{V. H. Binh}
\email{vhbinh@iop.vast.vn}
\affiliation{Institute of Physics, Vietnam Academy of Science and Technology, 10 Dao Tan,
	Ba Dinh, Hanoi, Vietnam}
\author{Cesar Bonilla}
\affiliation{Departamento de F\'{\i}sica, Universidad Cat\'olica del Norte, Avenida
	Angamos 0610, Casilla 1280, Antofagasta, Chile}
\email{cesar.bonilla@ucn.cl}
\author{A. E. C\'arcamo Hern\'andez}
\email{antonio.carcamo@usm.cl}
\affiliation{Universidad T\'ecnica Federico Santa Mar\'{\i}a, Casilla 110-V, Valpara\'{\i}%
	so, Chile}
\affiliation{Centro Cient\'{\i}fico-Tecnol\'ogico de Valpara\'{\i}so, Casilla 110-V,
	Valpara\'{\i}so, Chile}
\affiliation{Millennium Institute for Subatomic Physics at High-Energy Frontier (SAPHIR),
	Fern\'andez Concha 700, Santiago, Chile}
\author{D. T. Huong}
\email{dthuong@iop.vast.vn}
\affiliation{Institute of Physics, Vietnam Academy of Science and Technology, 10 Dao Tan,
	Ba Dinh, Hanoi, Vietnam}
\author{Vishnudath K. N.}
\email{vishnudath.neelakand@usm.cl}
\affiliation{Universidad T\'ecnica Federico Santa Mar\'{\i}a, Casilla 110-V, Valpara\'{\i}%
	so, Chile} 
\author{H. N. Long}
\email{hoangngoclong@vlu.edu.vn}
\affiliation{Subatomic Physics Research Group,
	Science and Technology Advanced Institute,\\
	Van Lang University, Ho Chi Minh City, Vietnam
}
\affiliation{ Faculty of Applied Technology, School of  Technology,  Van Lang University, Ho Chi Minh City, Vietnam
}
\author{P. N. Thu}
\email{thupn@utb.edu.vn}
\affiliation{Graduate University of Science and Technology, Vietnam Academy of Science and Technology, 18 Hoang Quoc Viet, Cau Giay, Hanoi, 10000, Vietnam}
\affiliation{Faculty of Natural Sciences - Technology,
	Tay Bac University, Quyet Tam Ward, Son La City, Son La 360000, Vietnam}

\author{Iv\'an Schmidt}
\footnote{Our friend and collaborator Iv\'an Schmidt passed away during the completion of this work. He will be sorely missed.}
\affiliation{Universidad T\'ecnica Federico Santa Mar\'{\i}a, Casilla 110-V, Valpara\'{\i}%
	so, Chile}
\affiliation{Centro Cient\'{\i}fico-Tecnol\'ogico de Valpara\'{\i}so, Casilla 110-V,
	Valpara\'{\i}so, Chile}
\affiliation{Millennium Institute for Subatomic Physics at High-Energy Frontier (SAPHIR),
	Fern\'andez Concha 700, Santiago, Chile}

\begin{abstract}

We propose two models based on the $SU(3)_C \times SU(3)_L \times U(1)_X$ gauge symmetry, each 
incorporating distinct inverse seesaw mechanisms for generating neutrino masses at the radiative level. Therefore, neutrino masses are suppressed by the radiative nature of the mass generation mechanism, which occurs after the spontaneous breaking of the global lepton number symmetry. Both scenarios discussed here are characterized by the presence of vector-like charged leptons, which are involved in generating the masses of the Standard Model charged leptons. These additional vector-like fermions contribute to the anomalous magnetic moments of the electron and the muon. We perform a detailed analysis of the scalar sectors, show that these models can successfully accommodate the observed baryon asymmetry through resonant leptogenesis, and compute charged lepton flavor-violating decays, such as $\mu \rightarrow e \gamma$. We discuss the constraints of the model arising from these processes and those associated the non-unitarity of the lepton mixing matrix.

\end{abstract}

\maketitle

%\footnote{Corresponding author. These authors contribute equaly to this work}%Either ABC or Huong as a first author.

\section{Introduction }

The Standard Model (SM) of particle physics is an extremely successful theory whose predictions are highly consistent with the data from various experiments. However, it still has a few limitations that motivate one to consider theories beyond the Standard Model (BSM). One of them is
the absence of a neutrino mass generation mechanism.
The existence of non-zero neutrino masses has been confirmed by several neutrino oscillation experiments~\cite{McDonald:2016ixn,Kajita:2016cak,deSalas:2020pgw}. Furthermore, the SM neither explains the observed dark matter relic abundance nor the matter-antimatter asymmetry of the Universe~\cite{Planck:2018vyg}. In addition, the SM also fails to explain the features such as the existence of three generations of fermions, electric charge quantization, and the observed hierarchy among the masses of the charged fermions. All of these point towards the need for a new theory, of which the SM can be considered as a low-energy effective theory. 

%\VKN{Among the various extensions of the SM that are proposed in literature to address some of the above mentioned issues, the ones that are based on the $SU(3)_C\times SU(3)_L\times U(1)_X$ gauge group is of special significance~\cite{Valle:1983dk,Pisano:1991ee,Frampton:1992wt,Foot:1994ym,Hoang:1995vq,CarcamoHernandez:2005ka,Chang:2006aa,Hernandez:2013mcf,CarcamoHernandez:2013krw,Boucenna:2014ela,CarcamoHernandez:2014wdl,Okada:2015bxa,Hernandez:2016eod,Fonseca:2016tbn,CarcamoHernandez:2017cwi,CarcamoHernandez:2018iel,CarcamoHernandez:2019vih,CarcamoHernandez:2019iwh,CarcamoHernandez:2019lhv,CarcamoHernandez:2020ehn,Binh:2020aal,CarcamoHernandez:2021tlv,Hernandez:2021zje,Thu:2023xai,Ciftci:2022lai,Ciftci:2022lrc}. 

Current theoretical frameworks are constructed with the aim of addressing some of the issues as mentioned above. One such framework involves extensions based on the $SU(3)_C\times SU(3)_L\times U(1)_X$ gauge group (known as 3-3-1 models). These models are of great interest because they explain the number of SM fermion families as well as the quantization of the electric charge~\cite{deSousaPires:1998jc,VanDong:2005ux}.
%VKN{These are often referred to as the 3-3-1 models and they provide an explanation for the number of SM fermion families as well as the quantization of the electric charge~\cite{deSousaPires:1998jc,VanDong:2005ux}.} 
These scenarios are characterized by providing additional sources of CP-violation 
%\Binh{They  also} have \VKN{additional} sources of CP violation 
\cite{Montero:1998yw,Montero:2005yb} as well as a natural Peccei-Quinn
symmetry, which solves the Strong-CP problem \cite{Pal:1994ba,Dias:2002gg,Dias:2003zt,Dias:2003iq}. In addition, the 3-3-1 models also predict that the weak mixing angle $\theta_W$ has the following upper bound,  $\sin\theta^2_W<\frac1{4}$. Moreover, incorporating heavy sterile neutrinos into the fermion spectrum of 3-3-1 models may facilitate the emergence of potential candidates for cold dark matter, such as weakly interacting massive particles (WIMPs)~\cite{Mizukoshi:2010ky,Dias:2010vt,Alvares:2012qv,Cogollo:2014jia}.
%Furthermore, the inclusion of heavy sterile neutrinos in the \Binh{spectrum of fermions in the 3-3-1 models might allow the appearence of }%fermionic spectrum of the 3-3-1 models allows to have cold dark matter candidates \Binh{like} weakly interacting massive particles (WIMPs) \cite{Mizukoshi:2010ky,Dias:2010vt,Alvares:2012qv,Cogollo:2014jia}. 
A concise review of WIMPs in 3-3-1 electroweak eauge models is given in Ref. \cite{daSilva:2014qba}.

In 3-3-1 models, with the addition of extra Majorana neutrino gauge singlets, one can also incorporate a low-scale testable version of the traditional type-I seesaw mechanism for neutrino mass generation~\cite{Mohapatra:1986bd,Akhmedov:1995ip,Akhmedov:1995vm,Malinsky:2005bi}. The advantage of the low-scale seesaw models such as the inverse or linear seesaw is that one can lower the scale of heavy particles without having to consider extremely small Yukawa couplings. This, in turn, makes these models testable in collider experiments. Typically, this is achieved by decoupling the scale of the seesaw from the scale of the lepton number violation. For instance, in the case of the inverse seesaw mechanism, the scale of lepton number violation usually corresponds to a small soft-breaking Majorana mass term for the exotic neutral fermions, which is $\sim \mathcal{O}$(keV), whereas the mass scale of the seesaw mediators is around $\sim \mathcal{O}$(TeV). Such models can lead to large unitarity violations of the leptonic mixing matrix that can be probed in neutrino oscillation experiments, as well sizeable rates for charged lepton flavor violating decays, such as $\mu \rightarrow e \gamma$. Moreover, the small mass splitting between the heavy pseudo-Dirac neutral leptons in these low-scale seesaw realizations facilitates a viable scenario for resonant leptogenesis~\cite{Pilaftsis:1997jf,Gu:2010xc,Dolan:2018qpy,Dib:2019jod,Blanchet:2009kk,Blanchet:2010kw,Hernandez:2021xet,Hernandez:2021kju,Hernandez:2021uxx,Abada:2023zbb}. Low scale seesaw in the context of 3-3-1 models have been studied earlier in~\cite{CarcamoHernandez:2013krw,CarcamoHernandez:2017cwi,CarcamoHernandez:2018iel,CarcamoHernandez:2019iwh,CarcamoHernandez:2019vih,CarcamoHernandez:2019lhv,CarcamoHernandez:2020ehn,CarcamoHernandez:2021tlv,Hernandez:2021zje}.

%Finally, suppose one considers 3-3-1 electroweak gauge models with three right-handed Majorana neutrinos and %without  \Binh{none of} exotic charges. In that case, one can implement a low-scale linear or inverse seesaw mechanism \Binh{which is} useful for \Binh{either} generating the tiny active neutrino masses %while at the same time  \Binh{or} providing a favored scenario for resonant leptogenesis \Binh{at the same time}. It is worth mentioning that in the inverse seesaw realizations, the mixing between active and sterile neutrinos is sizable and several orders of magnitude larger than in the type I seesaw realization, thus giving rise to large unitarity violations of the PMNS leptonic mixing matrix, which could be probed in neutrino oscillation experiments. %Furthermore, such  \Binh{Moreover, these} sizeable mixings yield charged lepton flavor-violating decays within the reach of experimental sensitivity, \Binh{while they are unobservable in type I seesaw models.} %a situation that is in contrast with type I seesaw models where such processes are very tiny and are therefore unobservable. Furthermore,  \Binh{Besides,} in the inverse seesaw realizations, the small mass splitting between the heavy pseudo-Dirac neutral leptons allows a successful scenario for resonant leptogenesis. 

%\Binh{The motivations of this work are arised from }%In this work, motivated by  the considerations mentioned above. 
In this work, we propose two models based on the $SU(3)_C\times SU(3)_L\times U(1)_X$ gauge symmetry, featuring distinct radiative inverse seesaw realization for the generation of the tiny neutrino masses. In both models, the Majorana mass submatrix associated with the breaking of the lepton number by two units is radiatively generated, thereby explaining the small neutrino masses. In the first model, the inverse seesaw mechanism is implemented at a one-loop level, whereas in the second one, the neutrino mass generation mechanism is realized at a two-loop level. It's worth noting that the one-loop inverse seesaw model proposed here is, as far as we know, the most economical realization within the context of 3-3-1 models found in the literature. 
%{\color{orange} Let's be sure about it. I prefer to say it this way than criticizing another paper.} 
Furthermore, the model exhibits a preserved $Z_2$ symmetry arising from the spontaneous breaking of the global lepton number symmetry, which guarantees the radiative nature of the inverse seesaw.
%\VKN{Further,} the preserved $Z_2$ symmetry that guarantees the radiative nature of the inverse seesaw is not imposed by hand but arises from the spontaneous breaking of the global lepton number symmetry.
In both models, the one- and two-loop realizations, the SM-charged fermion masses arise from a tree-level seesaw mechanism mediated by the exchange of heavy charged %vectors like 
vector-like leptons. These vector-like fermions also contribute to the anomalous magnetic moments of the muon and the electron, thus providing a natural link between the charged lepton mass generation mechanism and the $g-2$ anomalies. The layout of the remainder of the paper is as follows: Sec.\ref{model} discusses the symmetry and particle spectra of the two models, while their scalar sectors are studied in Sec.\ref{potential}.

The phenomenological consequences of these models for leptogenesis, charged lepton flavor violation, and the muon and electron anomalous magnetic moments are discussed in Sections \ref{lepto} and \ref{g-2}. Finally, we summarise and discuss the results in Sec.\ref{conclusion}. 

%\VKN{In this section, we discuss the particle spectrum and symmetries of the two radiative inverse seesaw models wherein the small lepton number violating Majorana mass submatrix is generated at one and two loops. 
 %Before delving into the specifics of these models, we examine the rationale behind incorporating additional scalars, fermions, and symmetries. These components serve dual purposes: firstly, to accommodate inverse seesaw mechanism at one and two-loop levels, crucial for generating the observed tiny active neutrino masses, and secondly, to facilitate the generation of the masses for the Standard Model (SM)-charged leptons via tree level seesaw mechanism}.
 %\\
\section{The models}
\label{model} 

In what follows we present two different setups based on the $SU\left( 3\right) _{C}\times SU\left( 3\right) _{L}\times U\left(
1\right) _{X}$ gauge symmetry where the neutrino masses are generated via an inverse seesaw mechanism at the loop level. In such a scenario, left-handed fermions are in triplet representations of $SU(3)_L$ %the theory, 
hence, anomaly gauge cancellation is guaranteed. For this reason, these models offer a good explanation for the number of SM fermion families in nature. Another interesting result related to anomaly cancellation in these frameworks is the quantization of electric charge~\cite{VanDong:2005ux}. 
\\

 On the other hand, lacking the mechanism chosen by nature to account for small neutrino masses, the inverse seesaw mechanism \cite{Mohapatra:1986bd,Akhmedov:1995ip,Akhmedov:1995vm,Malinsky:2009df,Malinsky:2005bi} turns out to be appealing given the rich phenomenology \cite{Abada:2014vea,Deppisch:2015qwa,Abada:2021yot,Hernandez:2021xet,Hernandez:2021uxx,Bonilla:2023wok,Bonilla:2023egs,Abada:2023zbb}. The neutrino mass matrix, in this case,  has the following structure
 
\begin{equation}
M_{\nu }=\left( 
\begin{array}{ccc}
0_{3\times 3} & m & 0_{3\times 3} \\ 
m^{T} & 0_{3\times 3} & M \\ 
0_{3\times 3} & M^{T} & \mu%
\end{array}%
\right).  \label{Mnufull}
\end{equation}%
in the basis $\left( \nu _{L},\nu _{R}^{C},N_{R}^{C}\right)$. There are nine neutral fermions, three active neutrinos $\nu _{iL}$, and six $\nu _{iR}$ and $N_{iR}$ (with $i=1,2,3$) sterile neutrinos. Hence, $M_\nu$ turns out to be a $9\times9$ matrix. 
Assuming that the submatrices in Eq.~(\ref{Mnufull}) satisfy the hierarchy \LH{ $\mu \ll m \ll M$ } the light neutrino mass matrix is
\begin{equation}
m_\nu= m (M^T)^{-1}\mu M^{-1} m^T.  \notag
\end{equation}
From the last equation, one can see that the smallness of the neutrino masses is regulated by the linear dependence of the parameter $\mu$. That is, if $\mu\to0$ then $m_\nu=0$, implying lepton number conservation. One can argue that $\mu$ has a radiative origin and that the submatrices $m$ and $M$ arise at the tree level. This could lead to an explanation of the hierarchy $\mu \ll m \ll M$, and hence the small neutrino masses.

\subsection{Model-1: 3-3-1 model with inverse seesaw at one-loop}\label{model1}

We consider a theory with the gauge symmetry $SU\left( 3\right) _{C}\times SU\left( 3\right) _{L}\times U\left( 1\right) _{X}$ supplemented by two global symmetries, a generalized global lepton number $%
U\left( 1\right) _{L_{g}}$  and a discrete $Z_2$ symmetry. We will consider the situation in which both, gauge and global, symmetry groups are broken after spontaneous symmetry breaking. %\VKN{The $U\left( 1\right) _{L_{g}} \times Z_2$ symmetry is spontaneously broken.} 
The scalar sector of
the model consists of three $SU\left( 3\right) _{L}$ scalar triplets $%
\chi $, $\eta $ and $\rho $, as well as %two
three electrically neutral gauge
singlet scalars $\varphi$, $\sigma$, $\xi$. The $SU\left( 3\right) _{L}$ scalar triplet $\chi $
triggers the spontaneous breaking of the $SU\left( 3\right) _{L}\times
U\left( 1\right) _{X}$ symmetry down to the SM electroweak gauge group,
which is %futher
further broken spontaneously  by the $SU\left( 3\right) _{L}$ scalar
triplets $\eta $ and $\rho $. 
\\

The fermion sector of the model under
consideration includes: the fermionic particles of the conventional 3-3-1
model with right-handed neutrinos, three charged vector-like leptons $E_{i}$ ($i=1,2,3$) and 
two additional neutral leptons $\Psi _{nR}$ ($n=1,2$), transforming as singlets under the $SU\left( 3\right) _{L}$. In this scenario, it is expected a mix between the charged vector-like leptons $E_{i}$ and the SM-charged leptons, thereby triggering an extended seesaw mechanism that generates the SM-charged lepton masses.
Additionally, these vector-like leptons $E_{i}$ induce contributions to the muon and electron anomalous magnetic moments at the one-loop level. We summarize in Tables \ref{scalars1}, \ref{quarks1}, and \ref{leptons1} the particle content along with the charge assignments under the symmetries of the model. 
\\

\begin{table}[h]%[tbp]
\begin{tabular}{|c|c|c|c|c|c|}
\hline
& $SU\left( 3\right) _{C}$ & $SU\left( 3\right) _{L}$ & $U\left( 1\right)
_{X}$ & $U\left( 1\right) _{L_{g}}$ & $Z_{2}$ \\ \hline
$\chi $ & $\mathbf{1}$ & $\mathbf{3}$ & $-\frac{1}{3}$ & $\frac{4}{3}$ & $-1$
\\ \hline
$\eta $ & $\mathbf{1}$ & $\mathbf{3}$ & $-\frac{1}{3}$ & $-\frac{2}{3}$ & $1$
\\ \hline
$\rho $ & $\mathbf{1}$ & $\mathbf{3}$ & $\frac{2}{3}$ & $-\frac{2}{3}$ & $1$
\\ \hline
$\varphi $ & $\mathbf{1}$ & $\mathbf{1}$ & $0$ & $-\frac{1}{2}$ & $1$ \\ 
\hline
$\sigma $ & $\mathbf{1}$ & $\mathbf{1}$ & $0$ & $1$ & $1$ \\ \hline
$\xi $ & $1$ & $1$ & $0$ & $0$ & $-1$ \\ \hline
\end{tabular}%
\caption{Charge assignments of the scalar fields under $SU\left( 3\right) _{C}\times SU\left(
3\right) _{L}\times U\left( 1\right) _{X}\times U\left( 1\right) _{L_{g}} \times Z_2$
for Model-1.}
\label{scalars1}
\end{table}

\begin{table}[h]%[tbp]
\begin{tabular}{|c|c|c|c|c|c|}
\hline
& $SU\left( 3\right) _{C}$ & $SU\left( 3\right) _{L}$ & $U\left( 1\right)
_{X}$ & $U\left( 1\right) _{L_{g}}$ & $Z_{2}$ \\ \hline
$Q_{nL}$ & $\mathbf{3}$ & $\overline{\mathbf{3}}$ & $0$ & $\frac{2}{3}$ & $1$
\\ \hline
$Q_{3L}$ & $\mathbf{3}$ & $\mathbf{3}$ & $\frac{1}{3}$ & $-\frac{2}{3}$ & $1$
\\ \hline
$u_{iR}$ & $\mathbf{3}$ & $\mathbf{1}$ & $\frac{2}{3}$ & $0$ & $1$ \\ \hline
$d_{iR}$ & $\mathbf{3}$ & $\mathbf{1}$ & $-\frac{1}{3}$ & $0$ & $1$ \\ \hline
$T_{R}$ & $\mathbf{3}$ & $\mathbf{1}$ & $\frac{2}{3}$ & $-2$ & $-1$ \\ \hline
$B_{nR}$ & $\mathbf{3}$ & $\mathbf{1}$ & $-\frac{1}{3}$ & $2$ & $-1$ \\ 
\hline
\end{tabular}%
\caption{Charge assignments of the quarks under $SU\left( 3\right) _{C}\times SU\left(
3\right) _{L}\times U\left( 1\right) _{X}\times U\left( 1\right) _{L_{g}} \times Z_2$
for Model-1.}
\label{quarks1}
\end{table}

\begin{table}[h]%[tbp]
\begin{tabular}{|c|c|c|c|c|c|}
\hline
& $SU\left( 3\right) _{C}$ & $SU\left( 3\right) _{L}$ & $U\left( 1\right)
_{X}$ & $U\left( 1\right) _{L_{g}}$ & $Z_{2}$ \\ \hline
$L_{iL}$ & $\mathbf{1}$ & $\mathbf{3}$ & $-\frac{1}{3}$ & $\frac{1}{3}$ & $%
-1 $ \\ \hline
$l_{iR}$ & $\mathbf{1}$ & $\mathbf{1}$ & $-1$ & $1$ & $1$ \\ \hline
$E_{iL}$ & $\mathbf{1}$ & $\mathbf{1}$ & $-1$ & $1$ & $-1$ \\ \hline
$E_{iR}$ & $\mathbf{1}$ & $\mathbf{1}$ & $-1$ & $1$ & $-1$ \\ \hline
$N_{iR}$ & $\mathbf{1}$ & $\mathbf{1}$ & $0$ & $-1$ & $1$ \\ \hline
$\Psi _{nR}$ & $\mathbf{1}$ & $\mathbf{1}$ & $0$ & $\frac{1}{2}$ & $1$ \\ 
\hline
\end{tabular}%
\caption{Charge assignments of the leptons under $SU\left( 3\right) _{C}\times SU\left(
3\right) _{L}\times U\left( 1\right) _{X}\times U\left( 1\right) _{L_{g}} \times Z_2$
for Model-1.}
\label{leptons1}
\end{table}

Notice that neutral leptons $\Psi _{nR}$ ($n=1,2$) induce a one-loop level Majorana mass term for the gauge singlet neutral fermions $N_{iR}$ (the $\mu$ term in Eq.~\ref{Mnufull}), thus giving rise to an inverse seesaw mechanism at one-loop level. The breaking of generalized global lepton number, $U\left( 1\right) _{L_{g}}$ symmetry, into a residual $\widetilde{Z}_{2}$-symmetry is also relevant for this mechanism to work. This is evident from the transformations of the fields under the preserved $\widetilde{Z}_{2}$ symmetry, defined as $\left(
-1\right) ^{6L}$, with $L$ being the global lepton number. This implies that the fields $N_{iR}$, $\varphi $, and $\Psi _{nR}$ carrying a non-trivial charge under $U\left( 1\right) _{L_{g}}$ transform as under the residual symmetry as  $\widetilde{Z}_{2}$-odd, whereas the remaining fields are $\widetilde{Z}_{2}$-even. See Appendix~\ref{mutermloops} for more details about the operators required to successfully generate $\mu$ at the loop level.\\

In this model, the lepton number is defined as \cite{Chang:2006aa}: 
\begin{equation}
	L=\frac{4}{\sqrt{3}}T_{8}+L_{g},
\end{equation}%
where $L_{g}$ corresponds to %a conserved charge associated with  the $% U(1)_{L_{g}}$ generalized  lepton number symmetry 
the generalized lepton number~\cite{CarcamoHernandez:2017cwi}.
As mentioned, the remnant preserved $\widetilde{Z}_{2}$ symmetry allows for the
implementation of a one-loop level inverse seesaw mechanism that produces
the tiny active neutrino masses.
The electric charge in the considered model is given by the expression: 
\begin{equation*}
Q=T_{3}-\frac{1}{\sqrt{3}}T_{8}+X,
\end{equation*}%
where $T_{3}$ and $T_{8}$ are the $SU(3)_{L}$ generators and $X$ is the $U(1)_{X}$
charge. 
\\

The fermion representations of the $SU\left( 3\right) _{L}$ group are organized as follows:
%The $SU\left( 3\right) _{L}$ fermionic antitriplets and triplets of the %model are 
\begin{equation}
Q_{1L}=%
\begin{pmatrix}
u_{1} \\ 
d_{1} \\ 
J_{1} \\ 
\end{pmatrix}%
_{L}\sim \left( \mathbf{3},\mathbf{3},\frac{1}{3}\right) ,\hspace{0.2cm}%
Q_{nL}=%
\begin{pmatrix}
d_{n} \\ 
-u_{n} \\ 
J_{n} \\ 
\end{pmatrix}%
_{L}\sim \left( \mathbf{3},\mathbf{\bar{3}},0\right) ,\hspace{0.2cm}L_{iL}=%
\begin{pmatrix}
\nu _{i} \\ 
l_{i} \\ 
\nu _{i}^{c} \\ 
\end{pmatrix}%
_{L}\sim \left( \mathbf{1},\mathbf{3},-\frac{1}{3}\right) .  \notag
\end{equation}%
The $SU\left( 3\right) _{L}$ scalar triplets of this model are represented
as: 
\begin{equation}
\rho =%
\begin{pmatrix}
\rho _{1}^{+} \\ 
\frac{1}{\sqrt{2}}(v_{\rho }+R_{\rho }+ iI%
_{\rho }) \\ 
\rho _{3}^{+}%
\end{pmatrix}%
,\hspace{1cm}\eta =%
\begin{pmatrix}
\frac{1}{\sqrt{2}}(v_{\eta }+R_{\eta }^{1}+ iI%
_{\eta }^{1}) \\ 
\eta _{2}^{-} \\ 
R_{\eta }^{3}+ iI_{\eta }^{3}%
\end{pmatrix}%
,\hspace{1cm}\chi =%
\begin{pmatrix}
R_{\chi }^{1}+ iI_{\chi }^{1} \\ 
\chi _{2}^{-} \\ 
\frac{1}{\sqrt{2}}(v_{\chi }+R_{\chi }^{3}+ iI%
_{\chi }^{3})%
\end{pmatrix}%
\label{scalartriplet}\end{equation}%
 and the singlet scalars introduced to this model are:
 	\be
 	\varphi = \frac{1}{\sqrt{2}} \left(R_\varphi + i I_\varphi\right), \hspace{1cm} \xi = \frac{1}{\sqrt {2}} \left(v_\xi + R_\xi \right), \hspace{1cm} \sigma = \frac{1}{\sqrt{2}} \left(v_\sigma + R_\sigma + i I_\sigma\right).
 	\ee
 The singlet scalar $\xi$ is a real field that acquires a VEV. Further, both $\varphi$ and $\sigma$  are complex fields %but just
 \VKN{among which only}  $\sigma$ gets a non-zero value of  VEV.

 With the above-specified particle content and symmetries, the following Yukawa terms emerge for the quark and lepton sectors:
\begin{eqnarray}
-\mathcal{L}_{Y}^{\left( q\right) } &=&\sum_{n=1}^{2}\sum_{i=1}^{3}\left(
y_{u}\right) _{ni}\overline{Q}_{nL}\rho ^{\ast }u_{iR}+\sum_{i=1}^{3}\left(
y_{u}\right) _{3i}\overline{Q}_{3L}\eta
u_{iR}\notag \\
&&+\sum_{n=1}^{2}\sum_{i=1}^{3}\left( y_{d}\right) _{ni}\overline{Q}%
_{nL}\eta ^{\ast }d_{iR}+\sum_{i=1}^{3}\left( y_{d}\right) _{3i}\overline{Q}%
_{3L}\rho d_{iR}  \notag \\
&&+\sum_{n=1}^{2}\sum_{k=1}^{2}\left( y_{B}\right) _{nk}\overline{Q}%
_{nL}\chi ^{\ast }B_{kR}+y_{T}\overline{Q}_{3L}\chi T_{R}+ H.c
\end{eqnarray}%
\begin{eqnarray}
-\mathcal{L}_{Y}^{\left( l\right) }
&=&\sum_{i=1}^{3}\sum_{j=1}^{3}y_{ij}^{\left( E\right) }\overline{L}%
_{iL}\rho E_{jR}+\sum_{i=1}^{3}\sum_{j=1}^{3}y_{ij}^{\left( l\right) }%
\overline{E}_{iL}\xi l_{jR}+\sum_{i=1}^{3}\sum_{j=1}^{3}\left( m_{E}\right)
_{ij}\overline{E}_{iL}E_{jR}  \notag \\
&&+\frac{1}{2}\sum_{i=1}^{3}\left( x_{\rho }^{\left( L\right) }\right) _{ij}\varepsilon
_{abc}\overline{L}_{iL}^{a}\left( L_{jL}^{C}\right) ^{b}\left( \rho ^{\ast
}\right) ^{c}+\sum_{i=1}^{3}\sum_{j=1}^{3}\left(y_N\right)_{ij}\overline{%
L}_{iL}\chi N_{jR}  \notag \\
&&+\sum_{i=1}^{3}\sum_{n=1}^{2}\left( y_{N}\right) _{in}\overline{N}%
_{iR}\Psi _{nR}^{C}\varphi +\sum_{n=1}^{2}\sum_{m=1}^{2}\left( y_{\Psi
}\right) _{nm}\overline{\Psi }_{nR}\sigma \Psi _{mR}^{C}+H.c  \label{lyl}
\end{eqnarray}

\begin{figure}[h]
\begin{center}
%\vspace{-1}
%\begin{tabular}{cc}
\includegraphics[width=0.5\textwidth]{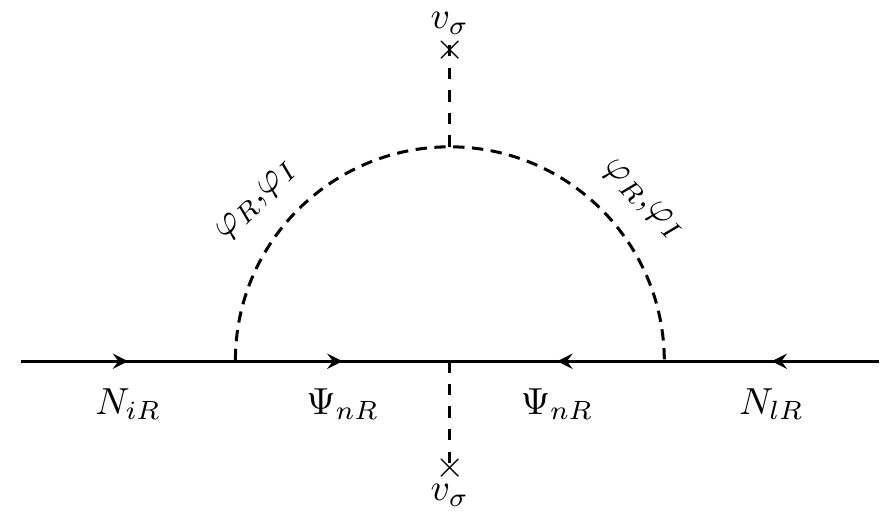} 
%\includegraphics[width=0.5\textwidth]{2loopmuv2.pdf}  
%\end{tabular}
%Here, $n,m=1,2$ and $i,j=1,2,3$.}
\end{center}%\vspace{1cm}
\caption{One-loop Feynman diagram contributing to the Majorana neutrino mass
submatrix $\protect\mu $ in model 1.} 
\label{Diagramsneutrinos1loop}
\end{figure}

The neutrino Yukawa interactions yield the following neutrino mass terms:
\begin{equation}
-\mathcal{L}_{mass}^{\left( \nu \right) }=\frac{1}{2}\left( 
\begin{array}{ccc}
\overline{\nu _{L}^{C}} & \overline{\nu _{R}} & \overline{N_{R}}%
\end{array}%
\right) M_{\nu }\left( 
\begin{array}{c}
\nu _{L} \\ 
\nu _{R}^{C} \\ 
N_{R}^{C}%
\end{array}%
\right)+\frac{1}{2}\sum_{n=1}^{2}\sum_{m=1}^{2}\left(m_{\Psi}\right)_{nm}\overline{\Psi }_{nR}\Psi _{mR}^{C} +H.c,  \label{Lnu}
\end{equation}%
where $m_{\Psi}=\sqrt{2}y_{\Psi} v_{\sigma}$ and the neutrino mass matrix $M_{\nu}$ has the form:
\begin{equation}
M_{\nu }=\left( 
\begin{array}{ccc}
0_{3\times 3} & m & 0_{3\times 3} \\ 
m^{T} & 0_{3\times 3} & M \\ 
0_{3\times 3} & M^{T} & \mu%
\end{array}%
\right) ,  \label{Mnu}
\end{equation}
with the neutrino submatrices $m$, $M$ and $\mu$ given by:
\begin{eqnarray}
m &=&\frac{v_{\rho }}{\sqrt{2}}\left[ \left( x_{\rho }^{\left( L\right)
}\right) ^{\dagger }-\left( x_{\rho }^{\left( L\right) }\right) ^{\ast }%
\right] ,\hspace{0.7cm}\hspace{0.7cm} M=\sqrt{2} y_{N} v_\chi, 
\notag \\
\mu _{sp} &=&\sum_{k=1}^{2}\frac{\left( y_{N}\right) _{sk}\left(
y_{N}^{T}\right) _{kp}m_{\Psi _{k}}}{16\pi ^{2}}\left[ \frac{m_{R
_{\varphi}}^{2}}{m_{R_{\varphi}}^{2}-m_{\Psi _{k}}^{2}}\ln \left( \frac{%
m_{R _{\varphi}}^{2}}{m_{\Psi _{k}}^{2}}\right) -\frac{m_{I_{\varphi}}^{2}}{%
m_{I_{\varphi}}^{2}-m_{\Psi _{k}}^{2}}\ln \left( \frac{m_{I_{\varphi}}^{2}%
}{m_{\Psi _{k}}^{2}}\right) \right]\,,
\end{eqnarray}
where the masses of the real and imaginary parts of $\varphi$ are given by the Eq.(\ref{mRvarphi}) and Eq.(\ref{mIvarphi}). The one-loop diagram that generates the $\mu$ term given above is shown in Fig.~\ref{Diagramsneutrinos1loop}.

%\subsection{\Binh{The 3-3-1} model with two loop level inverse seesaw \Binh{(model 2)} }

\subsection{Model-2: 3-3-1 model with inverse seesaw at two-loops}

Now we discuss our second %theory 
model, which is also based on the gauge %symmetry \VKN{group} 
$SU\left( 3\right)
_{C}\times SU\left( 3\right) _{L}\times U\left( 1\right) _{X}$  supplemented by the spontaneously broken  $U\left( 1\right) _{L_{g}}$ global lepton
number symmetry. %\textcolor{red}{ Similar to the previous model, both the $U\left( 1\right)_{L_{g}}$ and $Z_{2}$ symmetries are spontaneously broken - Antonio, please check this information}. Done. 
%Besides, % that,
As in the case of Model-1, the spontaneous breaking of the $U\left( 1\right) _{L_{g}}$ global lepton number
symmetry gives rise to a remnant $\widetilde{Z}_{2}$
symmetry. %The $Z_{2}^{(L_{g})}$ symmetry acts as a lepton number symmetry, where only leptons are charged, thereby forbidding proton decay. Furthermore, the 
This $\widetilde{Z}_{2}$ symmetry is assumed to be preserved %to prevent
and this prevents the
 inverse seesaw mechanism at the tree level. The charges of fields under the $\widetilde{Z}_{2}$ symmetry  are given as $\left( -1\right) ^{3L}$, where $L$ corresponds to
the $U\left( 1\right) _{L_{g}}$ charge of the particle under consideration.
In the proposed theory, the inverse seesaw mechanism generates the tiny
masses of the light-active neutrinos, implemented at a two-loop level.

 The scalar sector of the model under consideration is composed of  three $SU\left( 3\right) _{L}$ scalar triplets $\rho $, $\eta $ and $\chi $ 
	as given in Eq.(\ref{scalartriplet}) , four electrically charged scalar singlets $\zeta^{\pm }$, $\xi ^{\pm }$ and three singlet neutral complex scalars given below:
	\bea
		\sigma &=&\frac{1}{\sqrt{2}}\left( v_{\sigma }+R_{\sigma }+iI_{\sigma }\right)
		\,,\hspace{1cm}\phi =\frac{1}{\sqrt{2}}\left( v_{\phi }+R_{\phi }+iI_{\phi
		}\right) \,,\hspace{1cm}\varphi =\frac{1}{\sqrt{2}}\left( R_{\varphi
		}+iI_{\varphi }\right)\,. %\,,\hspace{1cm}\xi^\pm\,,\hspace{1cm}\zeta^\pm\,.
	\eea
	 Among the three singlet scalars, both $\phi$ and $\sigma$ acquire non-zero VEVs but $\varphi$ does not.
	\\
 The fermion content of the model remains the same as in the previous model where the inverse seesaw mechanism was implemented at one loop level. %In our proposed theory, the inclusion of the 
 As before, the spontaneously broken $U\left( 1\right) _{L_{g}}$
lepton number symmetry %is crucial. It 
serves to forbid the charged lepton, Yukawa
operator $\overline{L}_{iL}\rho l_{jR}$ %and to guarantee 
so that the SM charged
lepton masses are only generated from an extended seesaw-like mechanism
mediated by heavy charged vector-like leptons $E_{i}$ ($i=1,2,3$).
%Furthermore
 Moreover, the $U\left( 1\right) _{L_{g}}$ symmetry allows for the
implementation of the inverse seesaw mechanism at the two-loop level. 

The scalar, quark, and leptonic spectra of the model, along with their assignments under the symmetries, are shown in Tables \ref{scalars2}, \ref{quarks2}, and \ref{leptons2}, respectively.% \textcolor{green}{"}

\begin{table}[h]%[tbp]
\begin{tabular}{|c|c|c|c|c|}
\hline
& $SU\left( 3\right) _{C}$ & $SU\left( 3\right) _{L}$ & $U\left( 1\right)
_{X}$ & $U\left( 1\right) _{L_{g}}$ \\ \hline
$\chi $ & $\mathbf{1}$ & $\mathbf{3}$ & $-\frac{1}{3}$ & $\frac{4}{3}$ \\ 
\hline
$\eta $ & $\mathbf{1}$ & $\mathbf{3}$ & $-\frac{1}{3}$ & $-\frac{2}{3}$ \\ 
\hline
$\rho $ & $\mathbf{1}$ & $\mathbf{3}$ & $\frac{2}{3}$ & $-\frac{2}{3}$ \\ 
\hline
$\sigma $ & $\mathbf{1}$ & $\mathbf{1}$ & $0$ & $4$ \\ \hline
$\phi $ & $1$ & $1$ & $0$ & $10$ \\ \hline
$\varphi $ & $\mathbf{1}$ & $\mathbf{1}$ & $0$ & $1$ \\ \hline
$\zeta ^{\pm }$ & $\mathbf{1}$ & $\mathbf{1}$ & $\pm 1$ & $\mp 3$ \\ \hline
$\xi ^{\pm }$ & $1$ & $1$ & $\pm 1$ & $\pm 5$ \\ \hline
\end{tabular}%
\caption{Scalar assignments under $SU\left( 3\right) _{C}\times SU\left(
3\right) _{L}\times U\left( 1\right) _{X}\times U\left( 1\right) _{L_{g}}$
for model 2.}
\label{scalars2}
\end{table}
\begin{table}[h]%[tbp]
\begin{tabular}{|c|c|c|c|c|}
\hline
& $SU\left( 3\right) _{C}$ & $SU\left( 3\right) _{L}$ & $U\left( 1\right)
_{X}$ & $U\left( 1\right) _{L_{g}}$ \\ \hline
$Q_{nL}$ & $\mathbf{3}$ & $\overline{\mathbf{3}}$ & $0$ & $\frac{2}{3}$ \\ 
\hline
$Q_{3L}$ & $\mathbf{3}$ & $\mathbf{3}$ & $\frac{1}{3}$ & $-\frac{2}{3}$ \\ 
\hline
$u_{iR}$ & $\mathbf{3}$ & $\mathbf{1}$ & $\frac{2}{3}$ & $0$ \\ \hline
$d_{iR}$ & $\mathbf{3}$ & $\mathbf{1}$ & $-\frac{1}{3}$ & $0$ \\ \hline
$T_{R}$ & $\mathbf{3}$ & $\mathbf{1}$ & $\frac{2}{3}$ & $-2$ \\ \hline
$B_{nR}$ & $\mathbf{3}$ & $\mathbf{1}$ & $-\frac{1}{3}$ & $2$ \\ \hline
\end{tabular}%
\caption{Quark assignments under $SU\left( 3\right) _{C}\times SU\left(
3\right) _{L}\times U\left( 1\right) _{X}\times U\left( 1\right) _{L_{g}}$
for model 2.}
\label{quarks2}
\end{table}
\begin{table}[h]%[tbp]
\begin{tabular}{|c|c|c|c|c|}
\hline
& $SU\left( 3\right) _{C}$ & $SU\left( 3\right) _{L}$ & $U\left( 1\right)
_{X}$ & $U\left( 1\right) _{L_{g}}$ \\ \hline
$L_{iL}$ & $\mathbf{1}$ & $\mathbf{3}$ & $-\frac{1}{3}$ & $\frac{1}{3}$ \\ 
\hline
$l_{iR}$ & $\mathbf{1}$ & $\mathbf{1}$ & $1$ & $-5$ \\ \hline
$E_{iL}$ & $\mathbf{1}$ & $\mathbf{1}$ & $-1$ & $5$ \\ \hline
$E_{iR}$ & $\mathbf{1}$ & $\mathbf{1}$ & $-1$ & $1$ \\ \hline
$N_{iR}$ & $\mathbf{1}$ & $\mathbf{1}$ & $0$ & $-1$ \\ \hline
$\Psi _{nR}$ & $\mathbf{1}$ & $\mathbf{1}$ & $0$ & $0$ \\ \hline
\end{tabular}%
\caption{Lepton assignments under $SU\left( 3\right) _{C}\times SU\left(
3\right) _{L}\times U\left( 1\right) _{X}\times U\left( 1\right) _{L_{g}}$
for model 2.}
\label{leptons2}
\end{table}

\begin{figure}[h]
\begin{center}
%\vspace{-1}
%\begin{tabular}{cc}
\includegraphics[width=0.5\textwidth]{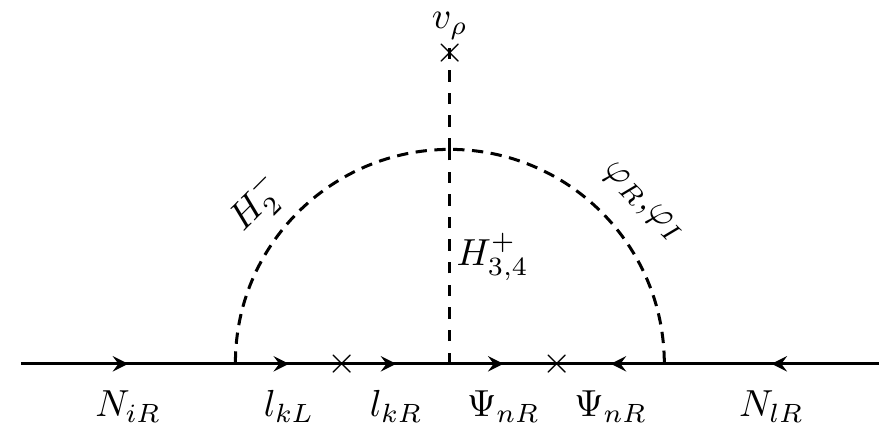} 
%\includegraphics[width=0.5\textwidth]{2loopmuv2.pdf}  
%\end{tabular}
%Here, $n,m=1,2$ and $i,j=1,2,3$.}
\end{center}%\vspace{1cm}
\caption{Two-loop Feynman diagram contributing to the Majorana neutrino mass
submatrix $\protect\mu $ in model 2.} 
\label{Diagramsneutrinos}
\end{figure}

With the above-specified particle content and symmetries, the following
Yukawa terms for the quark and lepton sectors arise: 
\begin{eqnarray}
-\mathcal{L}_{Y}^{\left( l\right) } &=&\sum_{i=1}^{3}\sum_{j=1}^{3}\left(
y_{E}\right) _{ij}\overline{L}_{iL}\rho
E_{jR}+\sum_{i=1}^{3}\sum_{j=1}^{3}\left( y_{l}\right) _{ij}\overline{E}%
_{iL}\phi l_{jR}+\sum_{i=1}^{3}\sum_{j=1}^{3}\left( x_{E}\right) _{ij}%
\overline{E}_{iL}\sigma E_{jR}  \notag \\
&&+\sum_{i=1}^{3}\sum_{j=1}^{3}\left( y_{N}\right) _{ij}\overline{L}%
_{iL}\chi N_{jR}+\sum_{i=1}^{3}\sum_{j=1}^{3}\left( y_{L}\right)
_{ij}\varepsilon _{abc}\overline{L}_{iL}^{a}\left( L_{jL}^{C}\right)
^{b}\left( \rho ^{\ast }\right) ^{c} \\
&&+\sum_{n=1}^{2}\sum_{k=1}^{2}\left( y_{\Psi }\right) _{nk}\overline{\Psi
_{kR}^{C}}\xi ^{+}l_{kR}+\sum_{n=1}^{2}\sum_{k=1}^{2}\left( x_{N}\right)
_{nk}\overline{N_{nR}^{C}}\varphi \Psi
_{kR}+\sum_{n=1}^{2}\sum_{k=1}^{2}\left( m_{\Psi }\right) _{nk}\Psi _{nR}%
\overline{\Psi _{kR}^{C}}+H.c  \label{lyl}
\end{eqnarray}

\begin{eqnarray}
-\mathcal{L}_{Y}^{\left( q\right) } &=&\sum_{n=1}^{2}\sum_{i=1}^{3}\left(
y_{u}\right) _{ni}\overline{Q}_{nL}\rho ^{\ast }u_{iR}+\sum_{i=1}^{3}\left(
y_{u}\right) _{3i}\overline{Q}_{3L}\eta
u_{iR}+\sum_{n=1}^{2}\sum_{i=1}^{3}\left( y_{d}\right) _{ni}\overline{Q}%
_{nL}\eta ^{\ast }d_{iR}+\sum_{i=1}^{3}\left( y_{d}\right) _{3i}\overline{Q}%
_{3L}\rho d_{iR}  \notag \\
&&+\sum_{n=1}^{2}\sum_{k=1}^{2}\left( y_{B}\right) _{nk}\overline{Q}%
_{nL}\chi ^{\ast }B_{kR}+y_{T}\overline{Q}_{3L}\chi T_{R}+H.c
\end{eqnarray}
The Majorana neutrino mass submatrix $\protect\mu $ is generated at two loops as shown in Fig.~\ref{Diagramsneutrinos} and takes the form \cite{Herrero-Garcia:2014hfa,McDonald:2003zj}:
\begin{eqnarray}
\mu &=&\sum_{k=3,4}\frac{\lambda _{\rho\varphi H_{2}^{+}H_{k}^{-}}v_{\rho }}{%
96\pi ^{2}m_{S}^{2}}\left( y_{N}\widetilde{M}_{l}g_{k}^{\dagger }m_{\Psi
}^{T}x_{N}^{T}+x_{N}m_{\Psi }g_{k}^{\ast }\widetilde{M}_{l}^{T}y_{N}^{T}%
\right) J\left( \frac{m_{H_{k}^{-}}^{2}}{m_{H_{2}^{+}}^{2}}\right) ,\end{eqnarray}
%\hspace{%1cm}
in which,
\begin{eqnarray}
    m_S&=&\max \left( m_{H _{k}^{-}},m_{H_2^{+}}\right) ,  \notag \\
J\left( \varkappa \right) &=&\left\{ 
\begin{array}{l}
1+\frac{3}{\pi ^{2}}\left( \ln ^{2}\varkappa -1\right) \hspace{0.1cm}\,
 \mbox{ for},\hspace{%
0.2cm}\varkappa \gg1 \\ 
\\ 
1\hspace{0.1cm}\,\mbox{for},\hspace{0.2cm}\varkappa \rightarrow 0.%
\end{array}%
\right. ,\end{eqnarray}
and the charged scalar fields $H_k^\pm$ $(k=3,4)$ are defined by the Eq.(\ref{Hcharged34}) with their masses are given by the Eq.(\ref{mHcharged34}) while 
%\hspace{1cm}\left( 
%\begin{array}{c}
%H_{3}^{\pm } \\ 
%H_{4}^{\pm }%
%\end{array}%
%\right) =\left( 
%\begin{array}{cc}
%\cos \theta & \sin \theta \\ 
%-\sin \theta & \cos \theta%
%\end{array}%
%\right) \left( 
%\begin{array}{c}
%\xi ^{\pm } \\ 
%\zeta ^{\pm }%
%\end{array}%
%\right) \\
\begin{eqnarray}
 g_{3} &=&y_{\Psi }\cos \alpha_{3_{(2)}} ,\hspace{1cm}g_{4}=-y_{\Psi }\sin \alpha_{3_{(2)}} .    
\end{eqnarray}

\section{Scalar potential and its tree level stability }
\label{potential} 
In this section, we %will 
examine the mass %spectrum
spectra of the scalar fields in detail for the two %versions mentioned
models discussed above. We divide the analysis into two parts, with each part corresponding to the investigation of one %version. 
variant.
%\subsection{\Binh{The 3-3-1} model with one loop level inverse seesaw}
\subsection{The scalar sector in Model-1}

\Binh{With} the scalar particle %structure 
\VKN{content} provided, one can construct \Binh{a scalar potential that is invariant under the} %transform of }%an invariant scalar potential with  the mentioned 
symmetries of the model as follows:
%The scalar potential of the model %is given by: 
\begin{eqnarray}
V_{1} &=&\mu _{\chi }^{2}\chi ^{\dag }\chi +\mu _{\rho }^{2}\rho ^{\dag
}\rho +\mu _{\eta }^{2}\eta ^{\dag }\eta +\mu _{\sigma }^{2}\sigma ^{\dag
}\sigma +\mu _{\varphi }^{2}\varphi ^{\dag }\varphi +\mu _{\xi }^{2}\xi
^{\dag }\xi + A(\varphi^2 \sigma +h.c)  \notag \\
&&+\lambda _{1}(\chi ^{\dag }\chi )^{2}+\lambda _{2}(\eta ^{\dag }\eta
)^{2}+\lambda _{3}(\rho ^{\dag }\rho )^{2}+\lambda _{4}(\sigma ^{\dag
}\sigma )^{2}+\lambda _{5}(\varphi ^{\dag }\varphi )^{2}+\lambda _{6}(\xi
^{\dag }\xi )^{2}  \notag \\
&&+\lambda _{7}(\chi ^{\dag }\chi )(\eta ^{\dag }\eta )+\lambda _{8}(\chi
^{\dag }\chi )(\rho ^{\dag }\rho )+\lambda _{9}(\eta ^{\dag }\eta )(\rho
^{\dag }\rho )+\lambda _{10}(\chi ^{\dag }\eta )(\eta ^{\dag }\chi )+\lambda
_{11}(\chi ^{\dag }\rho )(\rho ^{\dag }\chi )+\lambda _{12}(\eta ^{\dag
}\rho )(\rho ^{\dag }\eta )  \notag \\
&&+\lambda _{13}(\sigma ^{\dag }\sigma )(\chi ^{\dag }\chi )+\lambda
_{14}(\sigma ^{\dag }\sigma )(\rho ^{\dag }\rho )+\lambda _{15}(\sigma
^{\dag }\sigma )(\eta ^{\dag }\eta )+\lambda _{16}(\varphi ^{\dag }\varphi
)(\chi ^{\dag }\chi )+\lambda _{17}(\varphi ^{\dag }\varphi )(\rho ^{\dag
}\rho )+\lambda _{18}(\varphi ^{\dag }\varphi )(\eta ^{\dag }\eta )  \notag
\\
&&+\lambda _{19}(\xi ^{\dag }\xi )(\chi ^{\dag }\chi )+\lambda _{20}(\xi
^{\dag }\xi )(\rho ^{\dag }\rho )+\lambda _{21}(\xi ^{\dag }\xi )(\eta
^{\dag }\eta )+\lambda _{22}(\sigma ^{\dag }\sigma )(\varphi ^{\dag }\varphi
)+\lambda _{23}(\sigma ^{\dag }\sigma )(\xi ^{\dag }\xi )+\lambda
_{24}(\varphi ^{\dag }\varphi )(\xi ^{\dag }\xi )  \notag \\
&& + (B \epsilon^{ijk} \eta_i \rho_j \chi_k \xi + H.c) \,.
\label{potenScalar}
\end{eqnarray}

After expanding the scalar fields around their VEVs and substituting them into Eq. (\ref{potenScalar}), we can establish the following constraints at the tree level 
\begin{eqnarray}
\mu _{\chi }^{2}+\lambda _{1}v_{\chi }^{2}+\frac{1}{2}\left( \lambda
_{7}v_{\eta }^{2}+\lambda _{8}v_{\rho }^{2}+\lambda _{13}v_{\sigma
}^{2}+\lambda _{19}v_{\xi }^{2}\right) +B\frac{v_{\eta }v_{\rho }v_\xi}{%
2v_{\chi }} &=&0\,,  \notag \\
\mu _{\eta }^{2}+\lambda _{2}v_{\chi }^{2}+\frac{1}{2}\left( \lambda
_{7}v_{\chi }^{2}+\lambda _{9}v_{\rho }^{2}+\lambda _{15}v_{\sigma
}^{2}+\lambda _{21}v_{\xi }^{2}\right) +B\frac{v_{\chi }v_{\rho } v_\xi}{%
2v_{\eta }} &=&0\,,  \notag \\
\mu _{\rho }^{2}+\lambda _{3}v_{\chi }^{2}+\frac{1}{2}\left( \lambda
_{8}v_{\chi }^{2}+\lambda _{9}v_{\eta }^{2}+\lambda _{14}v_{\sigma
}^{2}+\lambda _{20}v_{\xi }^{2}\right) +B\frac{v_{\chi }v_{\eta }v_\xi}{%
2v_{\rho }} &=&0\,,  \notag \\
\mu _{\sigma }^{2}+\lambda _{4}v_{\sigma }^{2}+\frac{1}{2}\left( \lambda
_{13}v_{\chi }^{2}+\lambda _{14}v_{\rho }^{2}+\lambda _{15}v_{\eta
}^{2}+\lambda _{23}v_{\xi }^{2}\right) &=&0\,,  \notag \\
\mu _{\xi }^{2}+\lambda _{6}v_{\xi }^{2}+\frac{1}{2}\left( \lambda
_{19}v_{\chi }^{2}+\lambda _{20}v_{\rho }^{2}+\lambda _{21}v_{\eta
}^{2}+\lambda _{23}v_{\sigma }^{2}\right) +B\frac{v_{\eta }v_{\rho } v_\chi}{%
2v_{\xi }} &=&0.
\end{eqnarray}
\subsubsection{Charged scalar sector}
\label{potentialcc1} 
There are four charged scalar fields: $\rho_1^-,
\rho_3^-, \eta_2^-$, and $\chi_2^-$, which are mixed in pairs: \LH{$\rho_1^-$ mixes with $\eta_2^-$, and $\rho_3^-$ mixes with $\chi_2^-$.}
In the basis $(\rho_1^-, \eta_2^-)$, the \AC{squared mass} matrix is: 
\begin{eqnarray}
M^2_{c1_{(1)}} = \frac{\lambda _{12} v_\eta v_\rho-B v_\xi v_\chi}{2} \left( 
\begin{array}{cc}
\frac{v_\eta }{ v_\rho} & 1 \\ 
1 & \frac{v_\rho }{ v_\eta} \\ 
\end{array}
\right)\,.
\end{eqnarray}

 \LH{Note that here, the subscript $(1)$ stands for model 1 and later when we discuss model 2, we will use the subscript $(2)$. }
\LH{Diagonalizing this matrix, }we get one massless \LH{field}, which is identified to be the Goldstone boson $G_{1_{(1)}}^-$ eaten by $W^-_{(1)}$ boson and one massive boson $H_{1_{(1)}}^-$ with \AC{squared mass given by}:
	\begin{equation}
		m^2_{H_{1_{(1)}}^-} = \frac{\left(v_\eta^2+v_\rho^2\right)}{2 v_\eta v_\rho}%
		\left(\lambda _{12} v_\eta v_\rho-B v_\xi v_\chi\right)\,. \label{mH11pm}
	\end{equation}
 The physical fields are determined as:
\bea
G_{1_{(1)}}^- = \rho_1^- \cos \al_{1_{(1)}}  - \eta_2^- \sin \al_{1_{(1)}}\,,\\
H_{1_{(1)}}^-= \rho_1^- \sin \al_{1_{(1)}} +\eta_2^- \cos \al_{1_{(1)}}\,
\eea
where the mixing angle is given by the relation:
\be
\tan \al_{1_{(1)}} = - \frac{v_\eta}{v_\rho}\,.\label{al1}
\ee
%As the result of positive squared mass, the constraint of the parameter $B$ is:
\VKN{Ensuring the value of $m^2_{H_{1_{(1)}}^-}$ to be positive puts a constraint on the parameter $B$ as:}
\be B < \frac{\lambda _{12} v_\eta v_\rho}{v_\xi v_\chi}\,.
\ee

In the basis $(\rho_3^-, \chi_2^-)$, the \AC{squared mass matrix} is: 
\begin{eqnarray}
	M^2_{c2_{(1)}} = \frac{\lambda _{11} v_\chi v_\rho-B v_\xi v_\eta}{2} \left( 
	\begin{array}{cc}
		\frac{v_\chi }{ v_\rho} & 1 \\ 
		1 & \frac{v_\rho }{ v_\chi} \\ 
	\end{array}
	\right)\,. 
\end{eqnarray}
%On the physical basis,
\Binh{The physical states are:}
\bea
	G_{2_{(1)}}^- = \rho_3^- \cos \al_{2_{(1)}}  - \chi_2^- \sin \al_{2_{(1)}}\,,\\
	H_{2_{(1)}}^-= \rho_3^- \sin \al_{2_{(1)}} +\chi_2^- \cos \al_{2_{(1)}},
	\eea
\Binh{with the mixing angle $\al_{2_{(1)}}$is determined by
	\be
	\tan \al_{2_{(1)}} = - \frac{v_\rho}{v_\chi}\,.\label{al2}
	\ee 
 After diagonalizing, the mass-squared matrix  has the form given below:}
%the matrix $M^2_{c2_{(1)}} $ has a form as below:
\begin{eqnarray}
    M^2_{c2^{diag}_{(1)}} = \frac{\lambda _{11} v_\chi v_\rho-B v_\xi v_\eta}{2} \left( 
		\begin{array}{cc}
			0 & 0 \\ 
			0 & \frac{v_\rho^2 + v_\chi^2}{v_\rho v_\chi} \\ 
		\end{array}
		\right),\label{Mc21}
\end{eqnarray}
%and the physical states are defined 
The massless Goldstone boson $G_{2_{(1)}}^-$ \Binh{is} eaten by $Y^-_{(1)}$ boson \Binh{whereas the mass of $H_{2_{(1)}}^-$ is given by the non-zero eigenvalue of the matrix 
}
\VKN{Again, requiring the expression for the mass-squared of the field $H_{2_{(1)}}^-$ to be positive puts a constraint on the parameter $B$ as:}
	\be B < \frac{\lambda _{11} v_\chi v_\rho}{v_\xi v_\eta},
	\ee

\subsubsection{CP odd scalar sector}

%There are 
\VKN{The model contains} seven CP-odd scalar fields: $I_\sigma, I_\varphi, I_\chi^1, I_\chi^3, I_\rho, I_\eta^1, I_\eta^3$. \Binh{ Two of these, $I_\varphi$ and $I_\sigma$, are completely \LH{decoupled.} %independent, 
These themselves correspond to the physical states.  }
\Binh{The $I_{\sigma_{(1)}}$ field is massless whereas the \AC{squared} mass of the field $I_{\varphi_{(1)}}$ is given as:}
	\be m^2_{I_{\varphi_{(1)}}} = \sqrt{2} A v_\sigma+ \mu _\varphi ^2+\frac{1}{2} \left(\lambda _{16} v_\chi^2+ \lambda _{17} v_\rho^2+\lambda _{18} v_\eta^2+\lambda _{22}
	v_\sigma^2+\lambda _{24} v_\xi^2\right)\,.\label{mIvarphi}
	\ee
 \Binh{The mass in Eq.(\ref{mIvarphi}) gives the contribution to the neutrino submatrix $\mu$ as mentioned in the subsection \ref{model1}.} The field $I_{\sigma_{(1)}}$ can be identified as the Majoron, which is the Goldstone boson associated with the spontaneous breaking of the \LH{$U(1)_{L_g}$ global} lepton number symmetry. Despite being massless, this Goldstone is phenomenologically harmless since it is a gauge singlet. The Majoron can acquire a mass from a soft-breaking mass term in the scalar sector.

\Binh{The five fields left are mixed in two groups: $( I_\chi^3, I_\rho, I_\eta^1)$ and $( I_\chi^1, I_\eta^3)$. Note that $I_\chi^3, I_\rho$  and $ I_\eta^1$ are the components of the scalar representations whose VEVs are non-zero while $I_\chi^1$ and $ I_\eta^3$ are the components of the scalar representations with \LH{vanishing} VEVs.}\\
In the basis $( I_\chi^3, I_\rho, I_\eta^1)$, \VKN{the symmetric mass-squared matrix takes the following form:}
\begin{eqnarray}
M_{odd1_{(1)}}^2 = -\frac{B v_\xi}{2 } \left( 
\begin{array}{ccc}
\frac{v_\eta v_\rho}{ v_\chi} & v_\eta & v_\rho \\ 
  & \frac{v_\eta v_\chi}{ v_\rho} & v_\chi \\ 
  &   & \frac{v_\rho v_\chi}{ v_\eta}%
\end{array}
\right)\,.\end{eqnarray}
Applying the Euler method as used in \cite{Binh:2020aal}, the diagonalized matrix $M_{odd1^{diag}_{(1)}}^2$ takes the following form:
\begin{eqnarray}
M_{odd1^{diag}_{(1)}}^2= -\frac{B v_\xi}{2 } \left( 
\begin{array}{ccc}
	0 & 0 & 0 \\ 
	0 & 0 & 0 \\ 
	0 & 0 & \frac{v_\rho v_\chi}{v_\eta}+\frac{v_\chi v_\eta}{v_\rho}+
	\frac{v_\eta v_\rho}{v_\chi}
\end{array}
\right)\, .  \label{MCPodd11}
\end{eqnarray}\\
From Eq. \eqref{MCPodd11}, we see that the model has two massless Goldstone bosons $G_{Z^\prime_{(1)}}, G_{Z_{(1)}}$ and a massive field $A_{1_{(1)}}$
whose mass is given by the non-zero eigenvalue of the matrix  $M_{odd1_{(1)}}^2$.
The physical states ($G_{Z^\prime_{(1)}}, G_{Z_{(1)}}, A_{1_{(1)}}$) are related to the $( I_\chi^3, I_\rho, I_\eta^1)$ states as below:
\begin{eqnarray}
\left(\begin{array}{c} G_{Z^\prime_{(1)}} \\ G_{Z_{(1)}} \\A_{1_{(1)}} \end{array} \right)=\left(%
\begin{array}{ccc}
- \cos \alpha_{3_{(1)}} & - \sin \alpha_{1_{(1)}} \sin\alpha_{3_{(1)}} & \sin \alpha_{1_{(1)}} \cos \alpha_{3_{(1)}} \\ 
0 & \cos \alpha_{1_{(1)}} & \sin \alpha_{1_{(1)}} \\ 
\sin \alpha_{3_{(1)}} & - \sin \alpha_{1_{(1)}} \cos \alpha_{3_{(1)}} & \cos \alpha_{1_{(1)}} \cos\alpha_{3_{(1)}}%
\end{array}
\right)
\left(\begin{array}{c} I_\chi^3 \\ I_\rho \\ I_\eta^1 \end{array} \right)\,,  \label{Uodd1}
\end{eqnarray}
where the \Binh{third} mixing angle is defined by: 
\begin{eqnarray}
%\tan \alpha = -\frac{v_\eta}{v_\rho}\,, \hspace*{0.25cm} 
\tan \alpha_{3_{(1)}} =
- \frac{v_\eta}{v_{\chi} \sqrt{1+\frac{v_\eta^2}{v_\rho^2}}} = -\frac{v_\eta%
}{v_\chi}|\cos \alpha_{1_{(1)}}|\,.  \label{mixodd11}
\end{eqnarray}

In the basis $( I_\chi^1, I_\eta^3)$,  the mass-squared matrix is: 
\begin{eqnarray}
M^2_{odd2_{(1)}} =\frac{\left(B v_\xi v_\rho-\lambda _{10} v_\eta v_\chi
\right)}{2} \left( 
\begin{array}{cc}
\frac{v_\eta}{v_\chi} & 1 \\ 
1 & \frac{v_\chi}{v_\eta} \\ 
\end{array}
\right)\,.
\end{eqnarray}
In the %eigenstates 
\VKN{mass} basis, the matrix $M^2_{odd2_{(1)}}$ has the following form:
\begin{eqnarray}
    M^2_{odd2^{diag}_{(1)}}= \frac{\left(B v_\xi v_\rho-\lambda _{10} v_\eta v_\chi
		\right)}{2} \left( 
	\begin{array}{cc}
	0 & 0 \\ 
		0 & \frac{\left(v_\eta ^2+v_\chi^2\right) }{v_\eta v_\chi}\\ 
	\end{array}
	\right)\,.  \label{MCPodd12}
\end{eqnarray}
%This implies that  $M^2_{odd2_{(1)}}$ 
\VKN{Thus, the mixing of $I_\chi^1$ and $ I_\eta^3$ } provides one massless field, $%\Im 
G_{X^0_{(1)}}^\Im$, which is identified as the imaginary component of the Goldstone boson $G_{X^0_{(1)}}$ eaten by the bilepton gauge boson $X^{0}_{(1)}$, and one massive field, denoted as $A_{2_{(1)}}$. The physical fields are presented as follows:
\begin{align*}
%\Im 
G_{X^0_{(1)}}^\Im &= I_\chi^1 \cos \alpha_{4_{(1)}} - I_\eta^3 \sin \alpha_{4_{(1)}}, \,\,\,\,\,
A_{2_{(1)}} = I_\chi^1 \sin \alpha_{4_{(1)}} + I_\eta^3 \cos \alpha_{4_{(1)}}.
\end{align*}
The mixing angle is defined as
\be \tan \alpha_{4_{(1)}} = \frac{v_\eta}{v_\chi} \,.\label{al4}
\ee
Thus,  the CP odd sector consists of four massless fields ($ G_{Z_{(1)}}, G_{Z^\prime_{(1)}}, G_{X^0_{(1)}}^\Im$ and $I_{\sigma_{(1)}}$) and three massive fields  ($A_{1_{(1)}}, A_{2_{(1)}}$ and $I_{\varphi_{(1)}}$ ). The masses of the $A_1, A_2$ fields should be of the same scale as that of $v_\chi$.
\subsubsection{CP-even scalar sector \label{CPeven}}
 \VKN{The model consists of eight CP-even scalar fields: $R_\varphi, R_\sigma,
	R_\chi^1, R_\chi^3,R_\xi, R_\rho, R_\eta^1, R_\eta^3$. %Among them,  five acquired non-zero values of VEVs are mixed, except for $R_\varphi, R_\chi^1, R_\eta^3$. 
 Five of them $(R_\sigma, R_\chi^3, R_\xi, R_\rho, R_\eta^1)$ with non-zero VEVs mix among themselves. The remaining three acquire no VEVs. but only $R_\chi^1, R_\eta^3$ mix with each other. The remaining scalar $R_\varphi$ %remains
 does not mix with any other scalars and its mass
 is given as}:
\be
m^2_{R_{\varphi_{(1)}}}= m^2_{I_{\varphi_{(1)}}}.\label{mRvarphi} %+2\sqrt{2} A v_\sigma,. 
\ee
 \VKN{Thus $R_\varphi$ corresponds to the gauge eignestate as well as the mass eigenstate, just like $I_\varphi$. }
 % are themselves physical states. 
 The masses  of $R_{\varphi_{(1)}}$ and $I_{\varphi_{(1)}}$ are the same and should be at the $v_\chi$ scale. \Binh{Thus,} they can combine to form a complex scalar field with a mass also depending on the parameter $A$ which is the coefficient of the $\varphi^2 \sigma$ term in the potential $V_1$.\\
 In the basis $(R^1_\chi, R^3_\eta)$, the mass mixing matrix $M^2_{even1_{(1)}}$ is as same as the matrix $M^2_{odd2_{(1)}}$ in Eq. \eqref{MCPodd12}. One obtains $ G_{X^0_{(1)}}^{\Re}$ corresponding to the real component of the Goldstone boson $G_{X^0_{(1)}}$, which is absorbed by the gauge boson $X^{0}_{(1)}$, and a massive boson \Binh{$\mathcal{H}_{R_{(1)}}$ with mass $m^2_{\mathcal{H}_{R_{(1)}}} = m^2_{A_{2_{(1)}}}$}. %. They also combine into  a complex scalar field }. 
\Binh{ The mixing angle to define the physical fields is also $\alpha_{4_{(1)}}$ as in the Eq.(\ref{al4}).}

%\Binh{In the basis with five components}, 
 \VKN{The mass-squared matrix corresponding to the remaining five CP-even scalars $(R_\sigma, R_\chi^3, R_\xi, R_\rho, R_\eta^1)$ 
  is given as follows:}
%The second group includes $(R_\sigma, R_\chi^3, R_\xi, R_\rho, R_\eta^1)$; their squared mass matrix, which defines the mixing states, has the form:
\begin{eqnarray}
M^2_{even2_{(1)}}= \left( 
\begin{array}{ccccc}
2 \lambda _4 v_\sigma^2 & \lambda _{13} v_\sigma v_\chi & \lambda _{23}
v_\xi v_\sigma & \lambda _{14} v_\rho v_\sigma & \lambda _{15} v_\eta
v_\sigma \\ 
  & 2 \lambda _1 v_\chi^2-\frac{B v_\eta v_\xi
v_\rho}{2 v_\chi} & \frac{B v_\eta v_\rho}{2}+\lambda _{19} v_\xi v_\chi & 
\frac{B v_\eta v_\xi}{2}+\lambda _8 v_\rho v_\chi & \frac{B v_\xi v_\rho}{2}%
+\lambda _7 v_\eta v_\chi \\ 
  &   & 2 \lambda_6 v_\xi^2-\frac{B v_\eta v_\rho v_\chi}{2 v_\xi} & \frac{%
B v_\eta v_\chi}{2}+\lambda _{20} v_\xi v_\rho & \frac{B v_\rho v_\chi}{2}%
+\lambda _{21} v_\eta v_\xi \\ 
  &   &  & 2 \lambda _3
v_\rho^2-\frac{B v_\eta v_\xi v_\chi}{2 v_\rho} & \frac{B v_\xi v_\chi}{2}%
+\lambda _9 v_\eta v_\rho \\ 
  &   &   & & 2 \lambda _2 v_\eta^2-\frac{B
v_\xi v_\rho v_\chi}{2 v_\eta} \\  
\end{array}
\right)\,.  \label{MCPeven11}
\end{eqnarray}
We assume that $v_\sigma, v_\chi \gg v_\xi, v_\eta, v_\rho$. Due to this assumption, the matrix $M^2_{even2_{(1)}}$ can be separated into two blocks corresponding to the two bases  $(R_\sigma, R_\chi^3)$ and $(R_\xi, R_\rho, R_\eta^1)$. The first block \Binh{with two components} has larger VEVs, while the second one has smaller VEVs \Binh{($\sim$ EW scale)}. \\
In the basis $(R_\sigma, R_\chi^3)$, the squared mass matrix has the form:
\begin{eqnarray}
M^2_{even2a_{(1)}}= \left( 
\begin{array}{cc}
2 \lambda _4 v_\sigma^2 & \lambda _{13} v_\sigma v_\chi \\ 
\lambda _{13} v_\sigma v_\chi & 2 \lambda _1 v_\chi^2-\frac{B v_\xi v_\eta
v_\rho}{2 v_\chi} \\ 
\end{array}
\right)\,.  \label{MCPeven1a}
\end{eqnarray}
The matrix $M^2_{even2a_{(1)}}$  provides two heavy massive physical fields $ \mathcal{H}_{1_{(1)}}, \mathcal{H}_{2_{(1)}}$, which are related to the gauge eigenstates $(R_\sigma, R_\chi^3)$ as follows: 
\begin{eqnarray}
\left( 
\begin{array}{c}\mathcal{H}_{1_{(1)}} \\  \mathcal{H}_{2_{(1)}} \end{array} \right)= \left( 
\begin{array}{cc}
\cos \beta_{1_{(1)}} & \sin \beta_{1_{(1)}} \\ 
-\sin \beta_{1_{(1)}} & \cos \beta_{1_{(1)}} \\  
\end{array}
\right)\left( 
\begin{array}{c} R_\sigma \\ R_\chi^3\end{array} \right)\,,  \label{Ueven2A}
\end{eqnarray}
with the mixing angle defined by:
\begin{equation}
\tan 2\beta_{1_{(1)}} = \frac{4 \lambda _{13} v_\sigma v_\chi^2}{B
v_\xi v_\eta v_\rho+4 \lambda _4 v_\sigma^2 v_\chi-4 \lambda _1 v_\chi^3}\,.
\end{equation}
The \AC{squared masses} of the two scalars $\mathcal{H}_{1_{(1)}}$ and $\mathcal{H}_{2_{(1)}}$ are:% respectively given as below: 
\begin{eqnarray}
m^2_{\mathcal{H}_{{1,2}_{(1)}}} &=& \lambda_{1} v_\chi^2 + \lambda_4 v_\sigma^2
- \frac{B v_\eta v_\rho v_\xi}{4 v_\chi} \pm \sqrt{\lambda_{13}^2 v_\chi^2
v_\sigma^2 +\left(\frac{B v_\eta v_\rho v_\xi}{4 v_\chi} +
\lambda_{4}v_\sigma^2-\lambda_{1} v_\chi^2\right)^2}.
\end{eqnarray}

In the basis $(R_\xi, R_\rho, R_\eta^1)$, the mass-squared matrix is:
\begin{eqnarray}
	M^2_{even2b_{(1)}}&=& \left( 
	\begin{array}{ccc}
		2 \lambda _6 v_\xi^2-\frac{B v_\eta v_\rho v_\chi}{2 v_\xi} & \frac{B v_\eta
			v_\chi}{2}+\lambda _{20} v_\xi v_\rho & \frac{B v_\rho \text{v$\chi 
				$}}{2}+\lambda _{21} v_\eta v_\xi \\ 
		  & 2 \lambda _3 v_\rho^2-\frac{B v_\eta v_\xi v_\chi}{2 v_\rho} & \frac{B v_\xi \text{v$%
				\chi $}}{2}+\lambda _9 v_\eta v_\rho \\ 
		  &   & 2 \lambda _2 v_\eta^2-\frac{B v_\xi v_\rho 	v_\chi}{2 v_\eta}  
	\end{array}\right)
\end{eqnarray}
%Given that the VEVs, $v_\rho, v_\xi,v_\chi$, %are at the electroweak scale, the model %forecasts the existence of two additional %light Higgs bosons alongside the 125 GeV Higgs %boson. 
In the %typical 
\VKN{usual} scenario, the physical states will be connected to the %initial states 
\VKN{gauge eigenstates} by a $3 \times 3$ unitary matrix \Binh{(see Appendix.\ref{Diag1} for details)} as below: %$R_{H_{(1)}}$, as follows%:\textcolor{green}{need the form of $R_{H_{(1)}}$ or  $R^{-1}_{H_{(1)}}= U_{even_{2b_{(1)}}}$}
\begin{equation}\left( 
\begin{array}{c}
\mathcal{H}_{\Binh{3}_{(1)}} \\ 
h_{(1)}\\
\mathcal{H}_{\Binh{4}_{(1)}}
\end{array}
\right)
 \simeq \left(
\begin{array}{ccc}
	\cos \beta_{3_{(1)}} & \sin \beta_{2_{(1)}}
	\sin \beta_{3_{(1)}} & -\cos \beta_{2_{(1)}} \sin \beta_{3_{(1)}} \\
	-\sin \beta_{3_{(1)}} & \sin \beta_{2_{(1)}} \cos \beta_{3_{(1)}} & -\cos
	\beta_{2_{(1)}} \cos \beta_{3_{(1)}} \\
	0 & \cos \beta_{2_{(1)}} & \sin \beta_{2_{(1)}} \\
\end{array}
\right)
\left( 
\begin{array}{c}
R_{\xi } \\ 
R_{\rho } \\ 
R_{\eta }^{1}%
\end{array}%
\right).
\end{equation}

\Binh{
Then, the particle $h$ corresponds to the SM Higgs-like boson and its \AC{squared mass} is given by:
\be
m_{h_{(1)}}^2 = \frac{-B v_\chi v_\xi}{2}\left(\frac{v_\rho}{v_\eta} + \frac{v_\eta}{v_\rho}\right) = \frac{-B v_\chi v_\xi v^2}{2 v_\rho v_\eta}\,,
\ee
where $v^2 = v_\rho^2 + v_\eta^2 = 246^2$ GeV$^2$ and the parameter B should be negative with the absolute value of $B v_\chi$ %should be 
\VKN{being} in the EW scale like $v_\xi, v_\rho$ and $ v_\eta$. Masses of the two other scalars, $\mathcal{H}_{{3,4}_{(1)}}$ are:
\be
m^2_{\mathcal{H}_{{3,4}_{(1)}}}= -\frac{B v_\chi v_\rho v_\eta}{4v v_\xi}\left( \sqrt{v^2+16v_\xi^2} \pm v\right)\,.
\ee
Since $B v_\chi$ is at the EW scale, the $\mathcal{H}_{4_{(1)}}$ boson could potentially be the light Higgs boson as announced by the CMS collaboration~\cite{ATLAS:2016neq}. On the other hand, the scalar $\mathcal{H}_{3_{(1)}}$ can have a mass at the TeV or subTeV scale.}\\

 %Two Higgs particles have been determined, each with masses situated on the electroweak scale. One of these particles
%The particle $h$ corresponds to the SM Higgs-like boson, while the other is potentially the light Higgs boson as announced by CMS \cite{ATLAS:2016neq}.}

%\subsection{\Binh{The} $3-3-1$ model with two-loop level for inverse seesaw mechanism}  %\textcolor{green}{Unify presentation with II.A, II.B, III.A}}
\subsection{The scalar sector in Model-2 }

%\Binh{Given the spectrum of the particles as the Sec.\ref{model},} the scalar potential of the 3-3-1 model with the two-loop level for the inverse seesaw mechanism is formulated as follows:
\VKN{From the particle spectrum defined in Sec.\ref{model}, the gauge invariant scalar potential for model 2 with two-loop inverse seesaw mechanism is given as:}
\begin{eqnarray}
V_{2} &=&\mu _{\chi }^{2}\chi ^{\dag }\chi +\mu _{\rho }^{2}\rho ^{\dag
}\rho +\mu _{\eta }^{2}\eta ^{\dag }\eta +\mu _{\sigma }^{2}\sigma ^{\dag
}\sigma +\mu _{\varphi }^{2}\varphi ^{\dag }\varphi +\mu _{\zeta }^{2}\zeta
^{+}\zeta ^{-}+\mu _{\xi }^{2}\xi ^{+}\xi ^{-}+\mu _{\phi }^{2}\phi ^{\ast
}\phi +\left( f\varepsilon ^{ijk}\eta _{i}\rho _{j}\chi _{k}+H.c\right) 
\notag \\
&&+\lambda _{1}(\chi ^{\dag }\chi )^{2}+\lambda _{2}(\eta ^{\dag }\eta
)^{2}+\lambda _{3}(\rho ^{\dag }\rho )^{2}+\lambda _{4}(\sigma ^{\dag
}\sigma )^{2}+\lambda _{5}(\varphi ^{\dag }\varphi )^{2}+\lambda _{6}(\zeta
^{+}\zeta ^{-})^{2}+\lambda _{7}(\xi ^{+}\xi ^{-})^{2}  \notag \\
&&+\lambda _{8}(\chi ^{\dag }\chi )(\eta ^{\dag }\eta )+\lambda _{9}(\chi
^{\dag }\chi )(\rho ^{\dag }\rho )+\lambda _{10}(\eta ^{\dag }\eta )(\rho
^{\dag }\rho )+\lambda _{11}(\chi ^{\dag }\eta )(\eta ^{\dag }\chi )+\lambda
_{12}(\chi ^{\dag }\rho )(\rho ^{\dag }\chi )+\lambda _{13}(\eta ^{\dag
}\rho )(\rho ^{\dag }\eta )  \notag \\
&&+\lambda _{14}(\sigma ^{\dag }\sigma )(\chi ^{\dag }\chi )+\lambda
_{15}(\sigma ^{\dag }\sigma )(\rho ^{\dag }\rho )+\lambda _{16}(\sigma
^{\dag }\sigma )(\eta ^{\dag }\eta )+\lambda _{17}(\varphi ^{\dag }\varphi
)(\chi ^{\dag }\chi )+\lambda _{18}(\varphi ^{\dag }\varphi )(\rho ^{\dag
}\rho )+\lambda _{19}(\varphi ^{\dag }\varphi )(\eta ^{\dag }\eta )  \notag
\\
&&+\lambda _{20}(\sigma ^{\dag }\sigma )(\varphi ^{\dag }\varphi )+\lambda
_{21}(\sigma ^{\dag }\sigma )(\zeta ^{+}\zeta ^{-})+\lambda _{22}(\zeta
^{+}\zeta ^{-})(\chi ^{\dag }\chi )+\lambda _{23}(\zeta ^{+}\zeta ^{-})(\rho
^{\dag }\rho )+\lambda _{24}(\zeta ^{+}\zeta ^{-})(\eta ^{\dag }\eta
)+\lambda _{25}(\varphi ^{\dag }\varphi )(\zeta ^{+}\zeta ^{-})  \notag \\
&&+\lambda _{26}(\sigma ^{\dag }\sigma )(\xi ^{+}\xi ^{-})+\lambda _{27}(\xi
^{+}\xi ^{-})(\chi ^{\dag }\chi )+\lambda _{28}(\xi ^{+}\xi ^{-})(\rho
^{\dag }\rho )+\lambda _{29}(\xi ^{+}\xi ^{-})(\eta ^{\dag }\eta )+\lambda
_{30}(\varphi ^{\dag }\varphi )(\xi ^{+}\xi ^{-})  \notag \\
&&+\,\lambda _{31}\left( \chi ^{\dag }\rho \zeta ^{-}\varphi ^{\ast
}+H.c\right) +\lambda _{32}\left( \zeta ^{+}\xi ^{-}\sigma ^{2}+H.c\right)
+\lambda _{33}(\zeta ^{+}\zeta ^{-})(\xi ^{+}\xi ^{-})+\lambda _{34}\left[
(\chi ^{\dag }\eta )\varphi ^{2}+H.c\right]  \notag \\
&&+\lambda _{35}(\phi ^{\ast }\phi )^{2}+\lambda _{36}(\phi ^{\ast }\phi
)(\chi ^{\dag }\chi )+\lambda _{37}(\phi ^{\ast }\phi )(\eta ^{\dag }\eta
)+\lambda _{38}(\phi ^{\ast }\phi )(\rho ^{\dag }\rho )+\lambda _{39}(\phi
^{\ast }\phi )(\sigma ^{\dag }\sigma )  \notag \\
&&+\lambda _{40}(\phi ^{\ast }\phi )(\varphi ^{\dag }\varphi )+\lambda
_{41}(\phi ^{\ast }\phi )(\zeta ^{+}\zeta ^{-})+\lambda _{42}(\phi ^{\ast
}\phi )(\xi ^{+}\xi ^{-}).
%+\mu _{sb}\left( \phi ^{2}+h.c\right)
\label{potenScalar2}
\end{eqnarray}%
%The three electrically neutral singlet scalar fields are represented as: 
%\begin{equation*}
%\sigma =\frac{1}{\sqrt{2}}\left( v_{\sigma }+R_{\sigma }+iI_{\sigma }\right)
%\,,\hspace{1cm}\phi =\frac{1}{\sqrt{2}}\left( v_{\phi }+R_{\phi }+iI_{\phi
%}\right) \,,\hspace{1cm}\varphi =\frac{1}{\sqrt{2}}\left( R_{\varphi
%}+iI_{\varphi }\right) \,,
%\end{equation*}%
\Binh{Expanding the scalar fields around their VEVs} \VKN{and substituting} %replacing them 
into Eq.(\ref{potenScalar2}), we %can define 
\VKN{obtain} the following constraints at tree level:
\bea
\mu _{\chi }^{2}+\lambda _{1}v_{\chi }^{2}+\frac{1}{2}\left( \lambda
_{8}v_{\eta }^{2}+\lambda _{9}v_{\rho }^{2}+\lambda _{14}v_{\sigma
}^{2}+\lambda _{36}v_{\phi }^{2}\right) +f\frac{v_{\eta }v_{\rho }}{%
	\sqrt{2}v_{\chi }}=0,\nn\\ 
\mu _{\eta }^{2}+\lambda _{2}v_{\eta }^{2}+\frac{1}{2}\left( \lambda
_{8}v_{\chi }^{2}+\lambda _{10}v_{\rho }^{2}+\lambda _{16}v_{\sigma
}^{2}+\lambda _{37}v_{\phi }^{2}\right) +f\frac{v_{\eta }v_{\rho }}{%
	\sqrt{2}v_{\eta }}=0,\nn\\ 
\mu _{\rho}^{2}+\lambda _{3}v_{\rho }^{2}+\frac{1}{2}\left( \lambda
_{9}v_{\chi }^{2}+\lambda _{10}v_{\eta }^{2}+\lambda _{15}v_{\sigma
}^{2}+\lambda _{38}v_{\phi }^{2}\right) +f\frac{v_{\eta }v_{\rho }}{%
	\sqrt{2}v_{\chi }}=0,\nn\\ 
\mu _{\sigma }^{2}+\lambda _{4}v_{\sigma }^{2}+\frac{1}{2}\left( \lambda
_{14}v_{\chi }^{2}+\lambda _{15}v_{\rho }^{2}+\lambda _{16}v_{\eta
}^{2}+\lambda _{39}v_{\phi }^{2}\right)=0, \nn\\ 
\mu _{\phi }^{2}+\lambda _{35}v_{\phi }^{2}+\frac{1}{2}\left( \lambda
_{36}v_{\chi }^{2}+\lambda _{37}v_{\eta }^{2}+\lambda _{38}v_{\rho
}^{2}+\lambda _{39}v_{\sigma }^{2}\right)=0.
% +2\mu_{sb}=0
\eea
\subsubsection{Charged scalar sector}

\label{potentialcc2} 
There are six charged scalar fields \VKN{in this variant of the 3-3-1 model. They are}: $\rho_1^-$,
$\rho_3^-$, $\eta_2^-$, $\chi_2^-$, $\xi^-$, and $\zeta^-$, %which 
\VKN{and they} mix in pairs.
$\rho_1^-$ mixes with $\eta_2^-$, $\rho_3^-$ mixes with $\chi_2^-$, and  $\xi^-$ mixes with $\zeta^-$. \\
In the basis $(\rho_1^-, \eta_2^-)$, the squared mass %mixing 
matrix has the form: 
\begin{eqnarray}
M^2_{c1_{(2)}} = \frac{\lambda _{13} v_\eta v_\rho\Binh{-}\sqrt{2}f v_\chi}{2} \left( 
\begin{array}{cc}
\frac{v_\eta }{ v_\rho} & 1 \\ 
1 & \frac{v_\rho }{ v_\eta} \\ 
\end{array}
\right)\,. \end{eqnarray}
%After being diagonalized, the matrix %$M^2_{c1_{(2)}}$ changes into following form:
%\begin{eqnarray}
%    M^2_{c1^{diag}_{(2)}}= \frac{\lambda _{13}% v_\eta v_\rho-\sqrt{2}f v_\chi}{2} \left( 
%	\begin{array}{cc}
%		0 & 0 \\ 
%		0 & \frac{v_\rho^2 + v_\eta^2}{v_\rho %v_\eta} \\ 
%	\end{array}
%	\right)\,. \label{2Mc1}
%\end{eqnarray}
%The physical states arising from the squared mass matrix $M^2_{c1^{diag}_{(2)}}$  are 
\Binh{Diagonalizing the matrix $M^2_{c1_{(2)}}$ gives the physical states as the massless Goldstone boson $G_{1_{(2)}}^-= \rho_1^- \cos \al_{1_{(2)}}  + \eta_2^- \sin \al_{1_{(2)}}$ eaten by the longitudinal components of the $W^-_{(2)}$ boson and the massive charged scalar
 $H_{1_{(2)}}^- = -\rho_1^- \sin \al_{1_{(2)}} +\eta_2^- \cos \al_{1_{(2)}}$}  with mass given by: 
\begin{equation}
m^2_{H_{1_{(2)}}^-} = \frac{v^2}{2 v_\eta v_\rho}%
\left(\lambda _{13} v_\eta v_\rho\Binh{-}\sqrt{2}f v_\chi \right)\,,  \label{2mH1pm}
\end{equation}
and the mixing angle \Binh{ $\alpha_{1_{(2)}}=\alpha_{1_{(1)}}$ is the same as the one defined in the Eq.(\ref{al1}).}\\
%leads to the physical states $G_{1_{(2)}}^{\pm}$ and $H_{1_{(2)}}^{\pm}$, which are defined similarly to $G_{1_{(1)}}^{\pm}$ and $H_{1_{(1)}}^{\pm}$ in model 1.
%\begin{equation}
%\tan \beta^\prime_1 = - \frac{v_\eta}{v_\rho}\,.
%\end{equation}
%From equation (\ref{2mH1pm}), \Binh{because of the positive squared mass}, we get the constraint of \Binh{parameter $f$ as below}
\VKN{The requirement of the mass-squared to be positive in Eq.(\ref{2mH1pm}) puts the following constraint on the parameter $f$}: 
\begin{equation}
f<   \frac{ \lambda_{13} v_\eta v_\rho}{\sqrt2 v_\chi}\,.
\end{equation}

In the basis $(\rho_3^-, \chi_2^-)$, the squared mass matrix is: 
	\begin{eqnarray}
		M^2_{c2_{(2)}} = \frac{\lambda _{12} v_\chi v_\rho-\sqrt{2}f v_\eta}{2} \left( 
		\begin{array}{cc}
			\frac{v_\chi }{ v_\rho} & 1 \\ 
			1 & \frac{v_\rho }{ v_\chi} \\ 
		\end{array}
		\right)\,.\end{eqnarray}
  After diagonalization, we obtain a massless Goldstone boson, $G_{2_{(2)}}^{-}$, which is absorbed by the $Y^{-}_{(2)}$ bilepton gauge boson, and a massive field, $H_{2_{(2)}}^{-}$, with mass given by:
\begin{equation}
m^2_{H_{2_{(2)}}^{-}} = \frac{\left(v_\chi^2 + v_\rho^2\right)}{2 v_\rho v_\chi}\left(\lambda_{12} v_\rho v_\chi - \sqrt{2}f v_\eta\right), \label{2mH2pm}
\end{equation}
where the mixing angle is \Binh{$\alpha_{2_{(2)}}=\alpha_{2_{(1)}}$ is the same as the one in defined in Eq.(\ref{al2})}.
%, leading to the physical states $G_{2_{(2)}}^{-}$ and $H_{2_{(2)}}^{-}$, which are defined \Binh{as same} as $G_{2_{(1)}}^{-}$ and $H_{2_{(1)}}^{-}$ in model 1. %defined via  \begin{equation}
%\tan \beta^\prime_2 = - \frac{v_\rho}{v_\chi}\,.
%\end{equation}
%From the equation (\ref{2mH2pm}), \Binh{using the positive condition of the squared mass,} we get the constraint of \Binh{parameter} $f$:
\VKN{Again, requiring the expression for the mass-squared defined in Eq.(\ref{2mH2pm}) to be greater than 0, we obtain the following restriction on $f$: }
\begin{equation}
f<\lambda_{12} \frac{ v_\chi v_\rho}{\sqrt{2}v_\eta}\,.
\end{equation}
 In the basis $(\xi^-, \zeta^-)$, the squared mass matrix is: 
	\bea
		M^2_{c3_{(2)}}= \frac{1}{2} \left(\begin{array}{cc}
		2 \mu_\xi^2 +\la_{26} v_\sigma^2 + \la_{27} v_\chi^2 + \la_{28} v_\rho^2 + \la_{29} v_\eta^2 +\la_{42} v_\phi^2 & \la_{32} v_\sigma^2\\
		\la_{32} v_\sigma^2 & 2 \mu_\zeta^2 +\la_{21} v_\sigma^2 + \la_{22} v_\chi^2 + \la_{23} v_\rho^2 + \la_{24} v_\eta^2 +\la_{41} v_\phi^2 
		\end{array}
		\right)\,.		
	\eea
	We can consistently \AC{consider a simplified benchmark scenario where} %assume that 
 $\la_{21} = \la_{26}, \la_{22} = \la_{27}, \la_{23} = \la_{28}, \la_{24} = \la_{29}, \la_{41} = \la_{42}$. Then we get two massive fields with masses:
	\be
	m_{H_{3,4_{(2)}}}^- = \frac{1}{2} \left(\mu_\zeta ^2+\mu_\xi ^2+\lambda _{26} v_\sigma^2+\lambda _{27} v_\chi^2+\lambda _{28} v_\rho^2+\lambda _{29} v_\eta^2+\lambda _{42} v_\phi^2\pm \sqrt{\left(\mu_\zeta ^2-\mu_\xi^2\right)^2 +\lambda _{32}^2 v_\sigma^4}\right)\,.\label{mHcharged34}
	\ee
	In this case, the mixing angle $\al_{3_{(2)}}$ is defined by:
	\bea
	\tan \al_{3_{(2)}} %&=& - \frac{ \sqrt{1-\frac{ (\mu_\zeta -\mu_\xi )	(\mu_\zeta +\mu_\xi )}{\mu_\zeta ^2+\sqrt{\left(\mu_\zeta ^2-\mu_\xi				^2\right)^2+\lambda _{32}^2 v_\sigma^4}-\mu_\xi ^2}} 		\sqrt{\left(-\mu_\zeta ^2+\sqrt{\left(\mu_\zeta ^2-\mu_\xi					^2\right)^2+\lambda _{32}^2 v_\sigma^4}+\mu_\xi				^2\right)^2+\lambda _{32}^2 v_\sigma^4}}{\sqrt2 \sqrt{\left(\mu			\zeta ^2-\mu_\xi ^2\right)^2+\lambda _{32}^2 v_\sigma^4}}\nn\\
	&=&  - \frac{ 	\sqrt{\left(\sqrt{\left(\mu_\zeta ^2-\mu_\xi^2\right)^2+\lambda _{32}^2 v_\sigma^4}+\mu_\xi^2-\mu_\zeta ^2\right)^2+\lambda _{32}^2 v_\sigma^4}}{\sqrt2 \sqrt{\mu_\zeta ^2-\mu_\xi ^2+\sqrt{\left(\mu_\zeta ^2-\mu_\xi
				^2\right)^2+\lambda _{32}^2v_\sigma^4}}\left(\left(\mu
			\zeta ^2-\mu_\xi ^2\right)^2+\lambda _{32}^2 v_\sigma^4\right)^{\frac 1 {4}}}\,.
	\eea
	Then the physical states of these two charged scalar fields are given as:
	\bea
	H_{3_{(2)}}^- &=& \xi^\pm \cos \al_{3_{(2)}} + \zeta^\pm \sin \al_{3_{(2)}}\,,\nn\\
		H_{4_{(2)}}^- &=& -\xi^\pm \sin \al_{3_{(2)}} + \zeta^\pm \cos \al_{3_{(2)}}\,.\label{Hcharged34}
	\eea
\Binh{These two charged Higgs give contribution to the Majorana neutrino mass submatrix $\mu$ as shown in the Fig.\ref{Diagramsneutrinos}.}

\subsubsection{CP-odd scalar sector}

The model consists of eight CP-odd scalar fields: $I_\varphi, I_\phi, I_\sigma, I_\chi^1, I_\chi^3, I_\rho,
I_\eta^1, I_\eta^3$. The fields $I_\phi$ and $I_\sigma$ %remain \Binh{separating}
\VKN{do not mix with other fields.} The $I_\sigma$ can be identified with the Majoron, as in the  \Binh{model 1}. %331 model with the one-loop level inverse seesaw.
The $I_\varphi$ is corresponds to a massive physical field with the mass:
\be
m^2_{I_{\varphi_{(2)}}} = \frac{1}{2} \left(2 \mu_ \varphi ^2+\lambda _{19} v_\eta^2+\lambda _{18} v_\rho^2+\lambda _{20} v_\sigma^2+\lambda _{17} v_\chi^2+\lambda _{40}
v_\phi^2\right)\,.
\ee 
The \Binh{five} remaining fields, $I_\chi^1, I_\chi^3, I_\rho, I_\eta^1, I_\eta^3$,
mix in two groups: $(I_\chi^3, I_\rho, I_\eta^1)$ and $(I^1_\chi, I^3_\eta)$. In the basis $( I_\chi^3, I_\rho, I_\eta^1)$, the squared mass matrix for the electrically neutral CP-odd scalars takes the form:

\begin{eqnarray}
M_{odd_{1_{(2)}}}^2 = -\frac{f}{\sqrt{2 }} \left( 
\begin{array}{ccc}
\frac{v_\eta v_\rho}{ v_\chi} & v_\eta & v_\rho \\ 
  & \frac{v_\eta v_\chi}{ v_\rho} & v_\chi \\ 
  &  & \frac{v_\rho v_\chi}{ v_\eta}%
\end{array}
\right)\,.  \label{2MCPodd1}
\end{eqnarray}

The matrix \Binh{in the Eq.}(\ref{2MCPodd1}) can be diagonalized by the $3 \times 3$ matrix: 
\begin{eqnarray}
U_{odd_{1_{(2)}}}=\left(%
\begin{array}{ccc}
		- \cos \alpha_{4_{(2)}} & - \sin \alpha_{1_{(2)}} \sin\alpha_{4_{(2)}} & \sin \alpha_{1_{(2)}} \cos \alpha_{4_{(2)}} \\ 
		0 & \cos \alpha_{1_{(2)}} & \sin \alpha_{1_{(2)}} \\ 
		\sin \alpha_{4_{(2)}} & - \sin \alpha_{1_{(2)}} \cos \alpha_{4_{(2)}} & \cos \alpha_{1_{(2)}} \cos\alpha_{4_{(2)}}%
\end{array}
\right)\,,  \label{2Uodd1}
\end{eqnarray}
where the mixing angles are defined by: 
\begin{eqnarray}
\tan\alpha_{1_{(2)}}= \tan\alpha_{1_{(1)}}, \hspace*{0.25cm} 
\tan \alpha_{4_{(2)}} = \tan\alpha_{3_{(1)}}\,.  \label{2mixodd1}
\end{eqnarray}
After diagonalizing, the matrix $M_{odd_{1_{(2)}}}^2$ \Binh{in the Eq.}(\ref{2MCPodd1}) has
form: 
\begin{eqnarray}
M_{odd_{1_{(2)_{diag}}}}^2= -\frac{f}{\sqrt{2} } \left( 
\begin{array}{ccc}
0 & 0 & 0 \\ 
0 & 0 & 0 \\ 
0 & 0 & \frac{v_\rho v_\chi}{v_\eta}+\frac{v_\chi v_\eta}{v_\rho}+\frac{%
	v_\eta v_\rho}{v_\chi} \\ 
&  & 
\end{array}
\right)\,  \label{2MCPodd1diag}
\end{eqnarray}
Thus, we obtain two massless physical states, namely the Goldstone bosons $G_{Z_{(2)}}$ and $G_{Z^\prime_{(2)}}$, which are absorbed by the \AC{longitudinal components of the $Z$ and $Z^\prime$ gauge} bosons, respectively. In addition, there is a massive pseudoscalar with a mass-squared given by $-\frac{f}{\sqrt{2}} \left(\frac{v_\rho v_\chi}{v_\eta}+\frac{v_\chi v_\eta}{v_\rho}+\frac{v_\eta v_\rho}{v_\chi}\right)$. \Binh{Since this expression has to be positive, the parameter $f$ should always be negative.}\\
%\VHB{we need to comment about this mass of $A_{1_{(2)}}$ (which is $\frac{-B v_\xi}{2}(...)$) and $A_{1_{(1)}}$ (which is $\frac{-f}{\sqrt2}(...)$) }\\
In the basis $( I_\chi^1, I_\eta^3)$, the squared mass matrix for these two
electrically neutral CP odd scalars has the form: 
\begin{eqnarray}
M^2_{odd_{2_{(2)}}} =\frac{\left(-\sqrt{2}f v_\rho+\lambda _{11} v_\eta v_\chi
	\right)}{2} \left( 
\begin{array}{cc}
\frac{v_\eta}{v_\chi} & 1 \\ 
1 & \frac{v_\chi}{v_\eta} \\
\end{array}
\right)\,.  \label{2MCPodd2}
\end{eqnarray}
The matrix $M^2_{odd_{2_{(2)}}}$ in Eq.(\ref{2MCPodd2}) is diagonalized by the $%
2 \times 2 $ matrix: 
\begin{eqnarray}
U_{odd_{2_{(2)}}} = \left( 
\begin{array}{cc}
\cos \al_{5_{(2)}} & \sin \al_{5_{(2)}} \\ 
-\sin \al_{5_{(2)}} & \cos \al_{5_{(2)}} \\ 
\end{array}
\right)\,,  \label{Uodd1ce}
\end{eqnarray}
in which, 
\begin{equation}
\tan \al_{5_{(2)}} = \frac{v_\eta}{v_\chi} = \tan \al_{4_{(1)}}\,.\label{al5}
\end{equation}
Thus, we get one massless physical field $ G_{X^0_{(2)}}^\Im$, which can be identified as the imaginary component of the Goldstone boson $G_{X^0_{(2)}} $. % \textcolor{green}{$G_{X^0_{(2)}}^\Im$ similar to model 1 }obsorted by bilepton gauge boson $X^0_{(2)}$. 
\Binh{Another physical field is} a massive field, $A_{2_{(2)}}$, \AC{with squared mass given as} $%
m^2_{A_{2_{(2)}}}= \frac{\left(v_\eta ^2+v_\chi^2\right) }{2v_\eta v_\chi} \left(\lambda _{11} v_\eta v_\chi -\sqrt{2}f v_\rho
\right)$. Requiring this to be positive, we get the following constraint on %$\lambda_{11}$ and 
\AC{the $f$ parameter}: 
\begin{equation}
f<\lambda_{11} \frac{ v_\eta v_\chi}{ \sqrt{2}v_\rho }\,.
\end{equation}
%\textit{In CP odd sector, there are three massless fields which two of them are Goldstone boson $G_{Z}, G_{Z^\prime}$ eaten by $Z$ and $Z^\prime$ boson. 	There are two massive fields which are pseudo scalar.}

\subsubsection{CP even scalar sector}

The CP even scalar sector has eight scalar fields: $R_\varphi, R_\sigma,
R_\chi^1, R_\chi^3,R_\phi, R_\rho, R_\eta^1, R_\eta^3$. The field $R_\varphi$ %is unmixed 
does not mix with the other fields and its mass is given as:
\be
m^2_{R_{\varphi_{(2)}}} = \frac{1}{2} \left(2 \mu \varphi ^2+\lambda _{19} v_\eta^2+\lambda _{18} v_\rho^2+\lambda _{20} v_\sigma^2+\lambda _{17} v_\chi^2+\lambda _{40}
	v_\phi^2\right)= m^2_{I_{\varphi_{(2)}}}\,.
\ee

The remaining seven fields left mix in two groups: $(R_\sigma, R_\chi^3,R_\xi, R_\rho, R_\eta^1)$ that acquire non-zero values of  VEV and $(R_\chi^1, R_\eta^3)$ with zero VEV.
In the basis $(R_\chi^1, R_\eta^3)$, the mass squared matrix $M^2_{even_{2_{(2)}}}$ has the same form as the matrix $M^2_{odd_{1_{(2)}}}$ given in Eq.\eqref{2MCPodd2}. Thus, we get the massless field $ G_{X^0_{(2)}}^\Re$ which can be identifie to be the real component of the Goldstone boson which is eaten by the gauge boson $X^0_{(2)}$ and a massive boson $\mathcal{H}_{R_{(2)}}$ with mass $ m^2_{\mathcal{H}_{R_{(2)}}}= m^2_{A_{2_{(2)}}}$. \Binh{Note that} the mixing angle that relates the physical fields to the gauge eigenstates \Binh{in this case} is the same as  $\al_{5_{(2)}}$ \Binh{defined in Eq.(\ref{al5})}.\\
 
In the basis $(R_\sigma, R_\chi^3,R_\phi, R_\rho, R_\eta^1)$, the squared
mass matrix for these five electrically neutral CP even scalars has the form: 
\begin{eqnarray}
M^2_{even_{2_{(2)}}}= \left(
\begin{array}{ccccc}
	2 \lambda _4 v_\sigma^2 & \lambda _{14} v_\sigma v_\chi & \lambda _{39} v_\sigma v_\phi& \lambda _{15}
	v_\rho v_\sigma & \lambda _{16} v_\eta
	v_\sigma \\
	
	  & 2 \lambda _1 v_\chi^2-\frac{f v_\eta v_\rho}{\sqrt{2} v_\chi} &
	\lambda _{36} v_\chi v_\phi& \frac{f v_\eta}{\sqrt{2}}+\lambda _9 v_\rho v_\chi & \frac{f
		v_\rho}{\sqrt{2}}+\lambda _8 v_\eta v_\chi \\
	
	 &  & 2 \lambda _{35} v_\phi^2 & \lambda _{38}
	v_\rho v_\phi& \lambda _{37} v_\eta v_\phi \\
	
	  &   &  & 2 \lambda _3 v_\rho^2-\frac{f
		v_\eta v_\chi}{\sqrt{2} v_\rho} & \frac{f
		v_\chi}{\sqrt{2}}+\lambda _{10} v_\eta v_\rho \\
	
	  &   &  &   & 2 \lambda _2 v_\eta^2-\frac{f
		v_\rho v_\chi}{\sqrt{2} v_\eta} \\
\end{array}
\right). \label{2MCPeven1}
\end{eqnarray}
We assume that their VEVs follow the hierarchy $ v_\chi,v_\sigma \gg v_\phi,v_\eta, v_\rho$. Hence, the matrix $M^2_{even_{1_{(2)}}}$ can be separated into two blocks : the first block with higher VEVs and the second one with lower VEVs ($\sim$ EW scale).

In the basis $(R_\sigma, R_{\chi^3})$, the squared mass matrix has the following form:
\begin{eqnarray}
M^2_{even_{2a_{(2)}}}= \left(
\begin{array}{cc}
2 \lambda _4 v_\sigma^2 & \lambda _{14} v_\sigma
v_\chi \\
\lambda _{14} v_\sigma v_\chi & \Binh{-}\frac{f \text{v$_\eta
		$} v_\rho}{\sqrt{2} v_\chi}+2 \lambda _1
v_\chi^2 \\
\end{array}
\right)\,.  \label{2MCPeven1a}
\end{eqnarray}
%This matrix is diagonalized by the $2 \times 2 $ matrix below:
This matrix provides two physical states, $\mathcal{H}_{1_{(2)}}, \mathcal{H}_{2_{(2)}}$ which are related to the gauge eigenstates as below: 
\begin{eqnarray}
%U_{even_{2a_{(2)}}}
\left( 
\begin{array}{c} \mathcal{H}_{1_{(2)}} \\ \mathcal{H}_{2_{(2)}}
\end{array}
\right)= \left( 
\begin{array}{cc}
\cos \beta_{1_{(2)}} & \sin \beta_{1_{(2)}} \\ 
-\sin \beta_{1_{(2)}} & \cos \beta_{1_{(2)}} \\ 
\end{array}
\right) \left( 
\begin{array}{c} R_\sigma \\ R_\chi^3
\end{array}
\right) \,,  \label{2Ueven1A}
\end{eqnarray}
with the mixing angle defined as:
\begin{equation}
\tan 2\beta_{1_{(2)}} = -\frac{4 \lambda _{14} v_\sigma v_{\chi}^2}{\sqrt{2} f v_\eta v_\rho \Binh{+}4 \lambda _4
	v_\sigma^2 v_\chi \Binh{-}4 \lambda _1 v_{\chi}^3}.
\end{equation}
The masses of these two fields are given as
 \begin{eqnarray}
m^2_{\mathcal{H}_{1,2_{(2)}}} %&=& \lambda _4 v_\sigma^2+\lambda _1v_\chi^2  -\frac{f v_\eta  v_\rho}{2 \sqrt{2} v_\chi}  \pm\frac{\sqrt{f^2 v_\eta^2 v_\rho^2+4 \sqrt{2} f v_\eta	v_\rho v_\chi \left(\lambda _4 v_\sigma^2-\lambda_1 v_\chi^2\right)+8 v_\chi^2 \left(\left(\lambda _1	v_\chi^2-\lambda _4 v_\sigma^2\right){}^2+\lambda _{14}^2 v_\sigma^2 v_\chi^2\right)}}{2 \sqrt{2} v_\chi}\nn\\
&=& \lambda _4 v_\sigma^2+\lambda _1v_\chi^2  -\frac{f v_\eta  v_\rho}{2 \sqrt{2} v_\chi} \pm \sqrt{\left(\frac{f v_\eta v_\rho}{2\sqrt2 v_\chi}\right)^2 +\frac{f v_\eta v_\rho}{\sqrt2 v_\chi}\left(\lambda _4 v_\sigma^2-\lambda_1 v_\chi^2\right) + \left(\lambda _4 v_\sigma^2-\lambda_1 v_\chi^2\right)^2+\la_{14}^2 v_\sigma^2 v_\chi^2}\,.
\end{eqnarray}

In the basis $(R_\phi, R_\rho, R_\eta^1)$, the mass squared matrix is: 
\begin{eqnarray}
M^{2}_{even_{2b_{(2)}}}= \left(
\begin{array}{ccc}
2\lambda _{35}
v_\phi^2 & \lambda _{38} v_\rho v_{\phi} & \lambda _{37} v_\eta v_\phi \\

\lambda _{38} v_\rho v_\phi & \Binh{-}\frac{f v_\eta
	v_\chi}{\sqrt{2} v_\rho}+2 \lambda _3 v_{\rho}^2 & \lambda _{10} v_\eta v_\rho-\frac{f
	v_\chi}{\sqrt{2}} \\

\lambda _{37} v_\eta v_\phi & \lambda _{10}
v_\eta v_\rho-\frac{f v_\chi}{\sqrt{2}} &
-\frac{f v_\rho v_\chi}{\sqrt{2} v_\eta}+2
\lambda _2 v_\eta^2 \\
\end{array}
\right)\,.  \label{2MCPeven1b}
\end{eqnarray}
 The determinant of the matrix above being non-zero implies the existence of three physical states. In general, the physical states are related to the gauge eigenstates through a $3\times 3$ unitary matrix : %\textcolor{green}{need the form of $R_{H_{(2)}}$ or  $R^{-1}_{H_{(2)}}= U_{even_{2b_{(2)}}}$}
\begin{equation}
\Binh{\left( 
\begin{array}{c}
\mathcal{H}_{3_{(2)}} \\ 
h_{(2)}\\
\mathcal{H}_{4_{(2)}}
\end{array}
\right)
 \simeq \left(
\begin{array}{ccc}
	\cos \beta_{3_{(2)}} & \sin \beta_{2_{(2)}}
	\sin \beta_{3_{(2)}} & -\cos \beta_{2_{(2)}} \sin \beta_{3_{(2)}} \\
	-\sin \beta_{3_{(2)}} & \sin \beta_{2_{(2)}} \cos \beta_{3_{(2)}} & -\cos
	\beta_{2_{(2)}} \cos \beta_{3_{(2)}} \\
	0 & \cos \beta_{2_{(2)}} & \sin \beta_{2_{(2)}} \\
\end{array}
\right)}
\left( 
\begin{array}{c}
R_{\phi } \\ 
R_{\rho } \\ 
R_{\eta }^{1}%
\end{array}%
\right).
\end{equation}
Assuming, \Binh{$v_\phi \gg v_\rho, v_\eta$}, we obtain the approximate values of the masses and the mixing angles in \Binh{Appendix.\ref{Diag2}.} We have identified two Higgs particles with masses at the EW scale. One \Binh{of these two particles} is identified as the SM Higgs-like boson, while the other may correspond to the light Higgs boson reported by the CMS collaboration~\cite{ATLAS:2016neq}.

\section{Leptogenesis \label{lepto}}

Due to the presence of the three pairs of pseudo-Dirac heavy neutrinos, the model can explain the observed baryon asymmetry of the universe via resonant leptogenesis~\cite{Pilaftsis:1997jf,Gu:2010xc,Dib:2019jod,Blanchet:2009kk,Blanchet:2010kw}. We focus on the one-loop model, as the analysis for the two-loop case will be very similar. The lightest pair of the heavy pseudo-Dirac neutrinos, ${\nu_R}_1^\pm$, with masses $M_1^\pm = ( M \pm \mu )_{11}$, decay to produce a lepton asymmetry that can be enhanced by the small mass splitting between ${\nu_R}_1^+$ and ${\nu_R}_1^-$. 

The CP asymmetry thus generated is given as~\cite{Gu:2010xc},
\be
\epsilon_{\pm} \simeq \frac{\textrm{Im}\Big[   \Big( (({x_\rho^{(L)}}_+)^\dagger {x_\rho^{(L)}}_-)^2 \Big)_{11}\Big]}{8 \pi A_\pm} \frac{r}{r^2 + \frac{\Gamma_\mp^2}{{M_1^\mp}^2}}
\ee
where ${x_\rho^{(L)}}_\pm = \frac{{x_\rho^{(L)}}}{\sqrt{2}} (1 \pm \frac{1}{4}M^{-1}\mu)$, $r = \frac{{M_1^+}^2 - {M_1^-}^2}{M_1^+ M_1^-}$, $A_\pm = {x_\rho^{(L)}}_\pm^\dag {x_\rho^{(L)}}_\pm$ and $\Gamma_\pm = \frac{A_\pm {M_1}_\pm}{8 \pi}$.

%%%%%%%%%%%%%%%%%%%%%%%%%%%%%%%%%%%%%%%%%%%%%%%%%%%%%%%%%%%
\begin{figure*}[h]
\begin{center}
\includegraphics[scale=0.7]{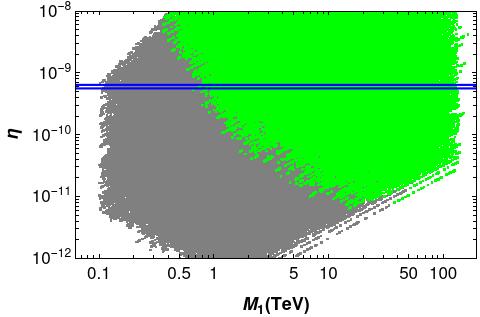} 
\includegraphics[scale=0.7]{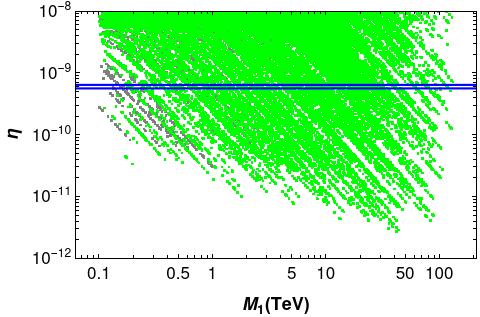}
\end{center}
\caption{ Baryon asymmetry parameter as a function of the mass of the lightest heavy neutrino. The figure in the left panel is for the parameter space with a weak washout, whereas the one in the right panel is for a strong washout. The horizontal blue lines correspond to the $3\sigma$ range of the observed baryon asymmetry of the universe. The gray points represent the correct active light neutrino masses, whereas the green points satisfy the bounds on the non-unitarity of the PMNS matrix.}
\label{figlep1}
\end{figure*}
%%%%%%%%%%%%%%%%%%%%%%%%%%%%%%%%%%%%%%%%%%%%%%%%%%%%%%%%%%%%%%%%
\begin{figure*}[h]
\begin{center}
\includegraphics[scale=0.7]{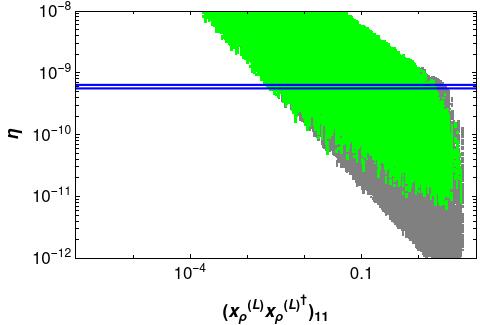} %
\includegraphics[scale=0.7]{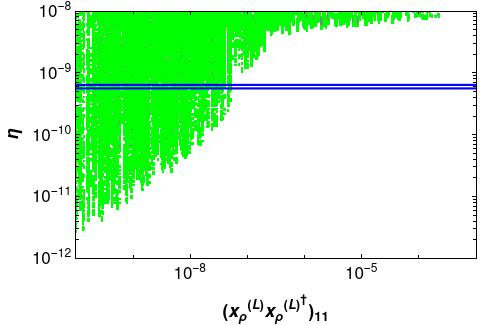}
\end{center}
\caption{Baryon asymmetry as a function of $((x_\rho^{(L)})^\dagger x_\rho^{(L)})_{11}$. The figure in the left panel is for the parameter space with a weak washout whereas the one in
the right panel is for a strong washout. The horizontal blue lines correspond
to the $3\sigma$ range of the observed baryon asymmetry of the universe. The color codes are the same as in Fig.\ref{figlep1}.}
\label{figlep2}
\end{figure*}
%%%%%%%%%%%%%%%%%%%%%%%%%%%%%%%%%%%%%%%%

Depending on the values of the couplings, leptogenesis can occur in either the strong washout regime or the weak washout regime. Assuming that the washout factor mainly depends on the inverse decay of the Higgs and charged lepton into the pseudo-Dirac neutrinos, the effective washout parameter is defined as,
\be K_N^{eff} \simeq \frac{\Gamma_+ + \Gamma_-}{H}  \Big(\frac{{M_1}_+ - {M_1}_-}{\Gamma_\pm}\Big)^2, \ee
with $H = \sqrt{\frac{4 \pi^3 g_* }{45}}\frac{T^2}{M_{pl}}$ is the Hubble parameter.
The expressions for the baryon asymmetry parameter in the weak and strong washout regimes are approximately given as\cite{Kolb:1990vq}:
\be \eta_B = \frac{\epsilon_\pm}{g_*}  ~~~~~ \textrm{for} ~~ K_N^{eff} \ll 1,\ee
\be \eta_B = \frac{0.3 \epsilon_\pm}{g_*  K_N^{eff} (\ln ~K_N^{eff})^{0.6}}  ~~~~~ \textrm{for} ~~ K_N^{eff} \gg 1.\ee

We emphasize that the model parameters influencing the baryon asymmetry, as defined above, also play a critical role in several other phenomena. These include the generation of light active neutrino masses, the non-unitarity of the neutrino mass mixing matrix, and processes involving lepton flavor violation. Fig. (\ref{figlep1}) displays the correlation of baryon asymmetry against the mass of the lightest heavy neutrino. The left panel represents the parameter space with a weak washout, while the right panel depicts a strong washout. The horizontal blue lines represent the $3\sigma$ range of the observed baryon asymmetry of the universe~\cite{Planck:2018vyg}. The gray points satisfy the bounds on active light neutrino masses and mixings as indicated by the oscillation experiments, whereas the green points satisfy the bounds on the non-unitarity of the PMNS matrix~\cite{Blennow:2023mqx}. Fig. (\ref{figlep2}) depicts the correlation between baryon asymmetry and $((x_\rho^{(L)})^\dagger x_\rho^{(L)})_{11}$. Here also, the left panel represents the parameter space with a weak washout, while the right panel illustrates a strong washout. All the color codes are the same as in Fig.~\ref{figlep1}. From these figures, we can see that the parameter space that successfully explains the observed matter-antimatter asymmetry in the weak washout regime corresponds to larger values of the Yukawa couplings whereas the parameter space with strong washout corresponds to very small Yukawa couplings. Because of this, the weak washout case does not get constrained by the bounds from the non-unitarity of the PMNS matrix as the unitarity violation is very small in this case. Further, in the weak washout case, the correct baryon asymmetry is satisfied for the lightest heavy neutrino mass $M_1 \gtrsim 1$ TeV, whereas in the strong washout regime, this is satisfied even for $M_1$ as light as a few $100$ GeV.
%%%%%%%%%%%%%%%%%%%%%%%%%%%
\begin{figure*}[h]
\begin{center}
\includegraphics[scale=0.55]{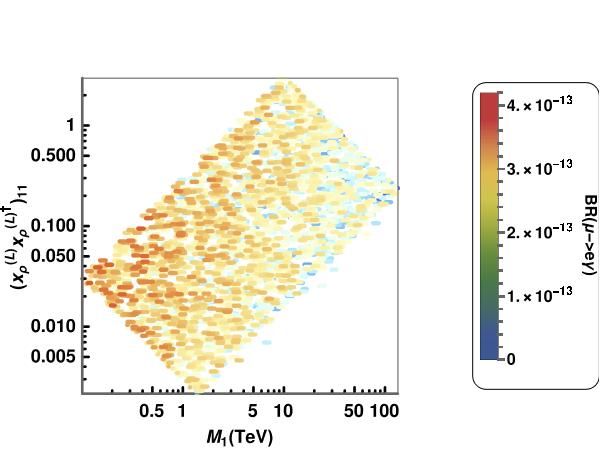} %
\includegraphics[scale=0.55]{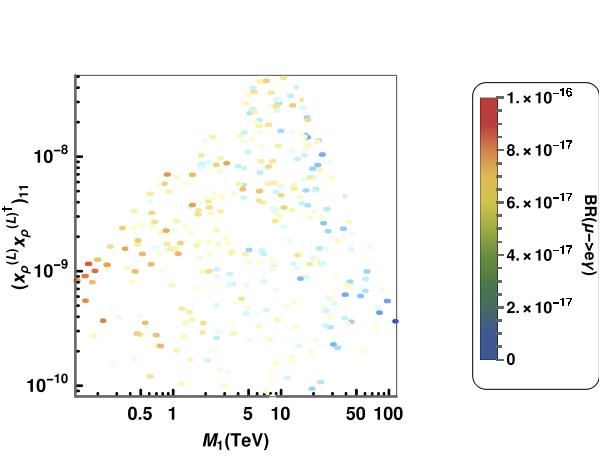}
\end{center}
\caption{Correlations of the mass of the lightest heavy neutrino against $((x_\rho^{(L)})^\dagger x_\rho^{(L)})_{11}$ for the parameter space that reproduces the correct value of the observed baryon asymmetry.  The figure in the left panel is for the parameter space with a weak washout whereas the one in
the right panel is for a strong washout. The color scales show the values of $BR(\mu \rightarrow e \gamma)$ for the corresponding points.  }
\label{figlep3}
\end{figure*}
%%%%%%%%%%%%%%%%%%%%%%%%%%%%%%%%%%%%%%%%%%%%%%%%%%%%%%%%%%%%
\begin{figure*}[h]
\begin{center}
\includegraphics[scale=0.7]{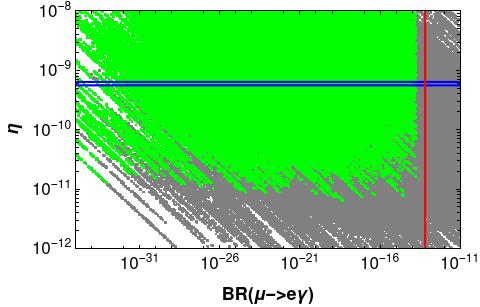} %
\includegraphics[scale=0.7]{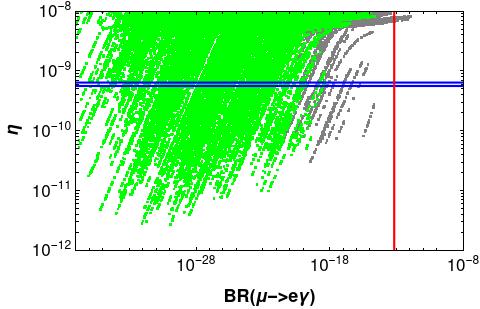}
\end{center}
\caption{Correlation of the baryon asymmetry asymmetry parameter against $BR(\mu \rightarrow e \gamma)$ for the scenarios of \AC{weak (left-panel) and strong (right-panel) washout}. The horizontal blue lines correspond to the $3\sigma$ range of the observed baryon asymmetry of the universe whereas the vertical red line corresponds to the MEG bound on $BR(\mu \rightarrow e \gamma)$. The color codes are the same as in Fig.\ref{figlep1}. }
\label{figlep4}
\end{figure*}

Fig. (\ref{figlep3}) illustrates the correlations between the mass of the lightest heavy neutrino and $((x_\rho^{(L)})^\dagger x_\rho^{(L)})_{11}$ for the parameter space that reproduces the correct value of the observed baryon asymmetry. The figures in the left and right panels correspond to the parameter spaces with weak washout and strong washout, respectively. The color scales represent the values of $BR(\mu \rightarrow e \gamma)$ for the corresponding points.  Fig. (\ref{figlep4}) illustrates the correlation of the baryon asymmetry parameter with $BR(\mu \rightarrow e \gamma)$. The horizontal blue lines represent the $3\sigma$ experimentally allowed range of the observed baryon asymmetry of the universe, while the vertical red line corresponds to the MEG bound on $BR(\mu \rightarrow e \gamma)$. Similarly, the color codes are consistent with those ones used in Figs. \ref{figlep1} and \ref{figlep2}. Once again, we can see that because of the small Yukawa couplings, the parameter space associated with the strong washout region predicts really small values for the branching ratio of the $\mu \rightarrow e \gamma$ decay, falling below the MEG limit. On the other hand, the weak washout case can give rise to larger values of $Br(\mu \rightarrow e \gamma)$ that could be observed in the future lepton flavor violating experiments. 

\section{Muon and electron anomalous magnetic moments\label{g-2}}
\AC{In this section we discuss the %consequences
\VKN{implications} of our models %in 
\VKN{for} the muon and electron anomalous magnetic moments. Given that the charged lepton sector is practically the same in both models (their differences %relies on 
\VKN{are mainly in the scalar and} neutrino sectors), the expressions for the muon and electron anomalous magnetic moments in the two models take the same form and are given by:} %defined by the following equations:
\begin{eqnarray}
\Delta a_{\mu } &\simeq &\dsum\limits_{k=1}^{3}\dsum\limits_{j=1}^{2}\frac{%
\func{Re}\left( \kappa _{2k}\gamma _{k2}^{\ast }\right) m_{\mu }^{2}}{8\pi
^{2}}\left( R_{H_{(1)}}\right) _{1j}\left( R_{H_{(1)}}\right) _{2j}I_{S}^{\left( \mu
\right) }\left( m_{E_{k}},m_{\mathcal{H}_{j}}\right) \notag \\&& +\dsum\limits_{k=1}^{3}\frac{%
\func{Re}\left( \kappa _{2k}\gamma _{k2}^{\ast }\right) m_{\mu }^{2}}{8\pi
^{2}}\left( R_{H_{(1)}}\right) _{13}\left( R_{H_{(1)}}\right) _{23}I_{S}^{\left( \mu
\right) }\left( m_{E_{k}},m_{h}\right), \label{Deltaamu}
\end{eqnarray}
\begin{eqnarray}
\Delta a_{e } &\simeq &\dsum\limits_{k=1}^{3}\dsum\limits_{j=1}^{2}\frac{%
\func{Re}\left( \kappa _{1k}\gamma _{k1}^{\ast }\right) m_{\mu }^{2}}{8\pi
^{2}}\left( R_{H_{(1)}}\right) _{1j}\left( R_{H_{(1)}}\right) _{2j}I_{S}^{\left( e
\right) }\left( m_{E_{k}},m_{\mathcal{H}_{j}}\right) \notag \\&& +\dsum\limits_{k=1}^{3}\frac{%
\func{Re}\left( \kappa _{1k}\gamma _{k1}^{\ast }\right) m_{\mu }^{2}}{8\pi
^{2}}\left( R_{H_{(1)}}\right) _{13}\left( R_{H_{(1)}}\right) _{23}I_{S}^{\left( e
\right) }\left( m_{E_{k}},m_{h}\right), % \notag \\
\label{Deltaae}
\end{eqnarray}

where, the loop function $I_{S\left( P\right) }^{\left( e,\mu \right)}\left( m_{E},m\right) $ has the form \cite%
{Diaz:2002uk,Jegerlehner:2009ry,Kelso:2014qka,Lindner:2016bgg,Kowalska:2017iqv}: 
\begin{equation}
I_{S\left( P\right) }^{\left( e,\mu \right) }\left( m_{E},m_{S}\right)
=\int_{0}^{1}\frac{x^{2}\left( 1-x\pm \frac{m_{E}}{m_{e,\mu }}\right) }{%
m_{\Binh{e,}\mu }^{2}x^{2}+\left( m_{E}^{2}-m_{e,\mu }^{2}\right) x+m_{S,P}^{2}\left(
1-x\right) }dx. \label{loopfunction}
\end{equation}

\VKN{In Eqs.(\ref{Deltaamu}) and (\ref{Deltaae}), $m_{E_k}$ correspond to the masses of exotic charged leptons $E_k$ and $m_{S,P}$ are the masses of scalar or pseudoscalar fields.}
\VKN{The matrix $R_{H_{(1)}}$ corresponds to the rotational matrix that diagonalizes the CP even scalar mass matrix. In general, it is a $3\times3$ orthogonal matrix.}

Besides, the dimensionless parameters $\beta _{1k}$, $\beta _{2k}$, $\gamma _{j1}$, and $\gamma _{j2}$ are given by:
\begin{eqnarray}
\kappa _{1j} &=&\dsum\limits_{i=1}^{3}y_{ij}^{\left( E\right) }\left(
V_{lL}^{\dagger }\right) _{1i},\hspace{0.7cm}\hspace{0.7cm}\gamma
_{j1}=\dsum\limits_{j=1}^{3}y_{kj}^{\left( l\right) }\left( V_{lR}\right)
_{j1}, \\
\kappa _{2j} &=&\dsum\limits_{i=1}^{3}y_{ij}^{\left( E\right) }\left(
V_{lL}^{\dagger }\right) _{2i},\hspace{0.7cm}\hspace{0.7cm}\gamma
_{j1}=\dsum\limits_{j=1}^{3}y_{kj}^{\left( l\right) }\left( V_{lR}\right)
_{j2},
\end{eqnarray}%
where $V_{lL}$ and $V_{lR}$ are the rotational matrices that diagonalize $%
\widetilde{M}_{l}$ according to the relation: 
\begin{equation}
V_{lL}^{\dagger }\widetilde{M}_{l}V_{lR}=\mathrm{diag}\left( m_{e},m_{\mu
},m_{\tau }\right) .
\end{equation}

\begin{figure}
    \centering
    \includegraphics[width=11.25cm, height=9cm]{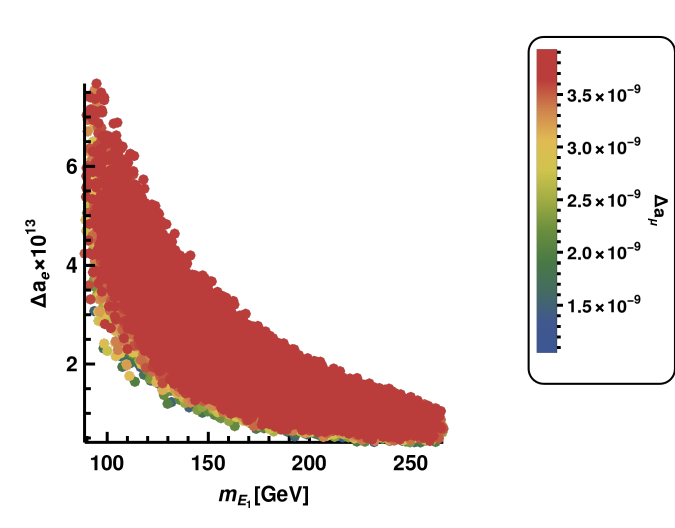}
    \caption{Corellation between $\Delta a_e$ and mass of charged lepton $E_1$.}
     %\caption{Corellations between $\Delta a_\mu$ and mass of charged lepton $E_2$.}
    \label{fig:eamm}
\end{figure}

\VKN{The experimentally measured values of the anomalous magnetic moments of the muon and electron are given as \cite{Muong-2:2023cdq,Morel:2020dww}: }
\begin{eqnarray}
\left( \Delta a_{\mu }\right) _{\exp } &=&\left(2.49\pm 0.48\right) \times
10^{-9}  \notag \\
(\Delta a_{e})_{\text{exp}} &=&(4.8\pm 3.0)\times 10^{-13}.
\end{eqnarray}
Since the value of the electron anomalous magnetic moment ($\Delta a_e$) is much smaller than that of the muon anomalous magnetic moment ($\Delta a_\mu$), one can focus solely on $\Delta a_e$ to establish reasonable correlations. %In Eq. (\ref{Deltaae}), the electron anomalous magnetic moment depends on the masses of the \Binh{exotic charged} %charged exotic  leptons $E_k$, where $k=1,2,3$. Thus, 
\VKN{In Fig.(\ref{fig:eamm})}, we plot the correlation of the electron anomalous magnetic moment $\Delta a_e$ \VKN{against} the mass of the \VKN{charged exotic} %charged exotic 
lepton $E_1$. %The Fig.(\ref{fig:eamm}) illustrates 
\VKN{The predictions for the muon anomalous magnetic moment are shown by the color scale. It can be seen from this figure that when the mass of $E_1$ varies from $100$ GeV to $250$ GeV, the prediction for  $\Delta a_\mu$ takes values up to a maximum of $3.5 \times 10^{-9}$ whereas the predictions for $\Delta a_e$ varies from $8\times 10^{-13}$ to $0.5 \times 10^{-13}$. Clearly, the value of the anomalous magnetic moment decreases with the increase in the mass of the charged exotic lepton.} 
\AC{As seen from Fig.~\ref{fig:eamm}, our proposed models are capable of successfully accommodating the experimental values of the muon and electron anomalous magnetic moments.}

\section{Conclusions \label{conclusion}}

\VKN{We have developed two models aimed at explaining the tiny masses of active light neutrinos. \AC{Both models} employ the inverse seesaw mechanism to generate small neutrino masses and \AC{the leptonic mixing angles}. In the first model, this is realized at the one-loop level, while in the second model, the inverse seesaw is ealized at the two-loop level. \AC{Both models} are extensions of the \AC{conventional 3-3-1 models}, where the gauge symmetry $SU(3)_C \times SU(3)_L \times U(1)_X$ is augmented by the inclusion of the generalized global lepton number symmetry $U(1)_{L_g}$. The spontaneous breaking of $U(1)_{L_g}$ to a preserved $Z_2$ symmetry ensures the radiative nature of the inverse seesaw mechanisms. Accomplishing this necessitates an extension of the fermion and scalar spectrum beyond the simplest versions of 3-3-1 models. Furthermore, in these models, the masses of the charged leptons in the SM arise from a tree-level seesaw mechanism.}

We have thoroughly examined the mass spectra of scalar particles in the two models. 

\VKN{Assuming that two of the scalars acquire VEVs of the order of EW scale, we note that there can be two CP-even scalar particles with masses in the range $100-200$ GeV. One of them corresponds to the SM Higgs boson of mass 125 GeV whereas the other could}
%The remaining particle may 
potentially be the Higgs boson with a mass of 150 GeV, as announced by \VKN{the} CMS \VKN{collaboration}. 

\VKN{We have also studied the parameter space for successful resonant leptogenesis focusing on the combined constraints from the light neutrino masses and mixings, non-unitarity of the PMNS matrix, and charged lepton flavor violation. We found that the model can simultaneously explain the observed baryon asymmetry and satisfy the experimental constraints, for both strong as well as weak washout regimes. We have also studied the contributions of the exotic vector-like charged leptons to the electron and muon anomalous magnetic moments and found that the model satisfies the experimental observations for charged exotic leptons in the mass range $100-250$ GeV.}

\section*{Acknowledgments}
RIP Iván, it was an honor to collaborate with you. The work of C.B. was supported by ANID-Chile FONDECYT Regular No. 1241855. AECH was funded by Chilean grants ANID-Chile FONDECYT 1210378, ANID PIA/APOYO AFB230003, ANID- Programa Milenio - code ICN2019\_044 and ANID-Chile FONDECYT Regular No. 1241855. V.K.N. is supported by ANID-Chile Fondecyt Postdoctoral grant 3220005. DTH acknowledges the financial support of the International Centre of Physics, Institute of Physics, VAST under Grant No.ICP-2024.02.

\appendix
\section{Radiative generation of the $\mu$-term}
\label{mutermloops}

\CB{
\subsection*{$\mu$-term at one-loop}
Here we give a list of conditions satisfied by Model-1 to have a $\mu$ matrix generated at the one-loop level. Fields and charge assignments in Tables~\ref{scalars1}, \ref{quarks1} and \ref{leptons1} allow the existence of following operators:%"
\begin{equation}
\overline{N}_{iR}\Psi _{nR}^{C}\varphi ,\hspace{1cm}\left( m_{\Psi }\right)
_{nk}\overline{\Psi }_{nR}\Psi _{kR}^{C},\hspace{1cm}\varphi ^{2}\sigma,\ \ (\text{with}\ \ i=1,2,3,\ \ \text{and}\ n,k=1,2)
\end{equation}
and forbid a tree-level operator:
\begin{equation}
\left( m_{N}\right) _{ij}\overline{N}_{iR}N_{jR}^{C},\ \ (\text{with}\ \ i,j=1,2,3),
\end{equation}
where $N_{iR}$ ($i=1,2,3$) and $\Psi _{nR}$ ($n=1,2$) are right-handed Majorana neutrinos, whereas $\varphi $ and $\sigma $ are electrically neutral-gauge singlet scalars. 
\\

The submatrices $m$ and $M$ are generated via the following operators:
\begin{equation}
\varepsilon _{abc}\overline{L}_{iL}^{a}\left( L_{jL}^{C}\right) ^{b}\left(
\rho ^{\ast }\right)^c ,\hspace{1cm}\overline{L}_{iL}\chi N_{jR},
\end{equation}
where $L_{iL}$ ($i=1,2,3$) and $\rho $ are $SU\left( 3\right) _{L}$ leptonic
and scalar triplets, respectively.
\\

Furthermore, the SM-charged lepton masses are generated from a tree-level seesaw mechanism \VKN{ with the charged lepton mass matrix in the basis %$\left( \overline{l}_{1L},\overline{l}_{3L},\overline{l}_{2L},\overline{E}_{1L}, \overline{E}_{2L},\overline{E}_{3L}\right) -\left( l_{1R},l_{2R},l_{3R},E_{1R},E_{2R},E_{3R}\right) $, 
 $(\overline{l}_{iL},\overline{E}_{iL}) - (\overline{l}_{iR},\overline{E}_{iR})$  
taking the following
structure: 
\begin{equation}
M_{l}=\left( 
\begin{array}{cc}
0_{3\times 3} & A \\ 
B^{T} & M_{E}%
\end{array}%
\right),
\end{equation}
where $i=1,2,3$, $E_{i}$  are charged exotic vector-like fermions and} $A$, $B$ and $M_{E}$ are $3\times 3$ matrices. Thus, the successful implementation of the tree-level seesaw mechanism to generate the SM-charged lepton masses requires that the operators: 
\begin{equation}
\overline{l}_{iL}\rho E_{jR},\hspace{1cm}\overline{E}_{iL}\xi l_{jR},\hspace{%
1cm}\left( M_{E}\right) _{ij}\overline{E}_{iL}E_{jR},
\end{equation}%
are forbidden whereas the operator: 
\begin{equation}
\overline{l}_{iL}\rho l_{jR},
\end{equation}%
is to be forbidden. This is guaranteed by discrete $Z_2$ symmetry.
\\

\VKN{Note that in addition to generating the SM-charged lepton masses via a seesaw-like mechanism, the inclusion of the exotic fermions $E_i$
also helps in successfully accommodating the experimental values of the muon and
electron anomalous magnetic moments.}

\subsection*{$\mu$-term at two-loops}
Now we proceed to discuss the considerations taken to generate the submatrix $\mu $
at the two-loop level in what we call Model-2. The particle content and charge assignments given in Tables~\ref{scalars2}, \ref{quarks2}, and \ref{leptons2}, guarantee the existence of the following operators:
\begin{eqnarray}
&&\overline{L}_{iL}\chi N_{jR},\hspace{1cm}\overline{\Psi _{kR}^{C}}\xi
^{+}l_{kR},\hspace{1cm}\overline{N_{nR}^{C}}\varphi \Psi _{kR},\hspace{1cm}%
\left( m_{\Psi }\right) _{nk}\Psi _{nR}\overline{\Psi _{kR}^{C}}, \\
&&\xi ^{-}\zeta ^{+}\sigma ^{2},\hspace{1cm}\chi ^{\dag }\rho \zeta
^{-}\varphi ^{\ast },
\end{eqnarray}%
and  forbid operators of the form:
\begin{equation}
\left( m_{N}\right) _{ij}\overline{N}_{iR}N_{jR}^{C},\hspace{1cm}\varphi
^{2}(\sigma ^{\dag }),\hspace{1cm}\varphi ^{2}\sigma ,\hspace{1cm}\varphi
^{2}(\sigma ^{\dag })^{2},\hspace{1cm}\varphi ^{2}\sigma ^{2},\hspace{1cm}%
i,j=1,2,3,
\end{equation}%
where $\sigma $ and $\varphi $ are electrically neutral scalar singlets \VKN{as before.}
 $\zeta ^{\pm }$ and $\xi ^{\pm }$\ are electrically charged scalars \VKN{whereas} $\eta $, $\chi $ and $\rho $ are $SU\left( 3\right) _{L}$
scalar triplets. %It is worth mentioning that  the implementation of  the radiative seesaw mechanism that generates
\VKN{Once again, the generation of} the submatrix $\mu $ at two-loop
level requires the inclusion of a spontaneously broken $U\left( 1\right)
_{L_{g}}$ symmetry %. Furthermore \Binh{Moreover}, it requires considering that the global lepton number symmetry $U\left( 1\right) _{L_{g}}$ 
\VKN{that} gets broken spontaneously
to a remnant preserved $\widetilde{Z}_{2}$ symmetry.
Under this $\widetilde{Z}_{2}$  symmetry, the scalar
field $\varphi $ has a non-zero %value of
charge and thus it does not acquire a
vacuum expectation value \Binh{(VEV)}. %The fact  \Binh{is} the global lepton number symmetry $U\left( 1\right) _{L_{g}}$ is spontaneously broken to a preserved remnant. 
\Binh{The} $%
\widetilde{Z}_{2}$ symmetry is crucial to forbid mass terms of the form $%
\left( m_{N\Psi }\right) _{nk}\overline{N_{nR}^{C}}\Psi _{kR}$ that would
result in a tree-level generation of the Majorana submatrix $\mu $.
Furthermore, in this realization of the two-loop level inverse seesaw
mechanism, the SM-charged lepton masses are generated via seesaw mechanism mediated by the following operators:
\begin{equation}
\overline{l}_{iL}\rho E_{jR},\hspace{1cm}\overline{E}_{iL}\phi l_{jR},%
\hspace{1cm}\overline{E}_{iL}\sigma E_{jR},
\end{equation}%

Similar to the situation in Model-1, the $U\left( 1\right) _{L_{g}}$ symmetry forbids the operator:
\begin{equation}
\overline{l}_{iL}\rho l_{jR},
\end{equation}
leading to a successful two-loop generation mechanism.
}

\section{Diagonalizing the mass matrix of the CP-even sectors in the model 1 }\label{Diag1}

In the basis $(R_\xi, R_\rho, R_\eta^1)$, the mass squared matrix is:
\begin{eqnarray}
	M^2_{even2b_{(1)}}&=& \left( 
	\begin{array}{ccc}
		2 \lambda _6 v_\xi^2-\frac{B v_\eta v_\rho v_\chi}{2 v_\xi} & \frac{B v_\eta
			v_\chi}{2}+\lambda _{20} v_\xi v_\rho & \frac{B v_\rho \text{v$\chi 
				$}}{2}+\lambda _{21} v_\eta v_\xi \\ 
		  & 2 \lambda _3 \text{v$%
			\rho $}^2-\frac{B v_\eta v_\xi v_\chi}{2 v_\rho} & \frac{B v_\xi \text{v$%
				\chi $}}{2}+\lambda _9 v_\eta v_\rho \\ 
		  &   & 2 \lambda _2 v_\eta^2-\frac{B v_\xi v_\rho 	v_\chi}{2 v_\eta} \\ 
	\end{array}
	\right)\nn\\
	&=& 2 v_\chi \left( 
	\begin{array}{ccc}
		\frac{ \lambda _6 v_\xi^2}{v_\chi}-\frac{B v_\eta v_\rho}{4 v_\xi} & \frac{B v_\eta	}{4}+ \frac{\lambda _{20} v_\xi v_\rho}{2v_\chi} & \frac{B v_\rho }{4}+ \frac{\lambda _{21} v_\eta v_\xi}{2v_\chi} \\ 
		& \frac{ \lambda _3 v_\rho^2}{v_\chi}-\frac{B v_\eta v_\xi }{4 v_\rho} & \frac{B v_\xi }{4}+ \frac{\lambda _9 v_\eta v_\rho}{2v_\chi} \\ 
		&   & \frac{ \lambda _2 v_\eta^2}{v_\chi}-\frac{B v_\xi v_\rho 	}{4 v_\eta} \\ 
	\end{array}
	\right)\,.  \label{MCPeven1b}
\end{eqnarray}
%Use 
\VKN{We take} the limit $v_\sigma, v_\chi \gg v_\xi, v_\eta, v_\rho$, \VKN{so that the terms that} 
%we can ignore some elements which 
are inversely proportional to $v_\chi$ in the matrix $	M^2_{even2b_{(1)}}$ \VKN{can be ignored}. Then, the approximate form of the matrix $	M^2_{even2b_{(1)}}$ becomes:
\begin{eqnarray}
	M^2_{even2B_{(1)}}&=& \frac{B v_\chi}{2} \left( 
	\begin{array}{ccc}
		%\frac{ \lambda _6 v_\xi^2}{v_\chi}
		-\frac{ v_\eta v_\rho}{ v_\xi} &  v_\eta%+ \frac{\lambda _{20} v_\xi v_\rho}{2v_\chi} 
		&  v_\rho%+ \frac{\lambda _{21} v_\eta v_\xi}{2v_\chi} 
		\\ 
		& %\frac{ \lambda _3 v_\rho^2}{v_\chi}
		-\frac{ v_\eta v_\xi }{ v_\rho} &  v_\xi %+ \frac{\lambda _9 v_\eta v_\rho}{2v_\chi} 
		\\ 
		&   & %\frac{ \lambda _2 v_\eta^2}{v_\chi}
		-\frac{ v_\xi v_\rho 	}{ v_\eta} \\ 
	\end{array}
	\right)= \frac{-B v_\chi }{2} M_{e2_{(1)}}\,. \label{MCPeven2B}
\end{eqnarray}
The $2 \times 2$ block on the right bottom of $		M^2_{even2B_{(1)}}$ may be rewritten as:
\begin{eqnarray}
	M^2_{e2_{(1)}}&=& \frac{-B v_\chi v_\xi}{2} \left( 
	\begin{array}{cc}
		\frac{v_\eta}{v_\rho} & 1 \\
		1 & \frac{v_\rho}{v_\eta}
	\end{array}
\right)\,.\end{eqnarray}
After diagonalizing, the matrix $M^2_{e2_{(1)}}$ has form:
\begin{eqnarray}
    M^2_{e2^{diag}_{(1)}}= \frac{-B v_\chi v_\xi}{2} \left( 
\begin{array}{cc}
	0 & 0 \\
	0 & \frac{v_\rho}{v_\eta} + \frac{v_\eta}{v_\rho}
\end{array}
\right)\,. \label{MCPeven2Bs}
\end{eqnarray}
Thus, we get a massive state $h_{(1)} = \cos \beta_{2_{(1)}} R_\rho +\sin \beta_{2_{(1)}} R_\eta^1 $. This could be the SM-like Higgs boson (SMLHB) whose mass is:
\be
m_{h_{(1)}}^2 = \frac{-B v_\chi v_\xi}{2}\left(\frac{v_\rho}{v_\eta} + \frac{v_\eta}{v_\rho}\right) = \frac{-B v_\chi v_\xi v^2}{2 v_\rho v_\eta}\,,
\ee
where $v^2 = v_\rho^2 + v_\eta^2 = 246^2$ GeV and the parameter B should be negative with the absolute value of $B v_\chi$ being in the EW scale as $v_\xi, v_\rho, v_\eta$. %about $\sim \frac{1}{v_\chi}$.
The matrix $	M_{e2_{(1)}}^2$  in Eq. \eqref{MCPeven2Bs} is diagonalized \VKN{by a $2\times2$ orthogonal matrix parametrized by} the mixing angle $\beta_{2_{(1)}}$ which is given by:
\be \tan \beta_{2_{(1)}} = \frac{v_\rho}{v_\eta} = - \cot \alpha_{1_{(1)}}\,.
\ee
Hence, the matrix that is used to diagonalize the matrix $	M_{e2_{(1)}}^2$ is:
\be
U_{\beta_{2_{(1)}}} = \left( 
\begin{array}{cc}
	\sin \beta_{2_{(1)}} & -\cos \beta_{2_{(1)}} \\ 
	\cos \beta_{2_{(1)}} & \sin \beta_{2_{(1)}} \\  
\end{array}
\right)\,.
\ee
\VKN{Coming to the diagonalization of the original $3\times 3$ mass squared matrix, the above $2\times 2$ can be represented as} %We receive 
a $3 \times 3$ rotational matrix with the following form:
\be
U_{\beta_{2s_{(1)}}} = \left( 
\begin{array}{ccc}
	1&0&0\\
	0&\sin \beta_{2_{(1)}} & -\cos \beta_{2_{(1)}} \\ 
	0&\cos \beta_{2_{(1)}} & \sin \beta_{2_{(1)}} \\  
\end{array}
\right)\,.
\ee
%%%%%%%%%%%%%%%%%%%%%%%%%%%%%%%%%%%%%%%%%%%%%%%%%%%%%%%%%%%%
This $U_{\beta_{2s_{(1)}}}$ matrix changes the matrix $M^2_{even2B_{(1)}}$ defined in B3 to another form:
\be
	M_{e2s_{(1)}} = -\frac{B v_\chi}{2}\left(
	\begin{array}{ccc}
		-\frac{v_\eta v_\rho}{v_\xi} & \frac{2 v_\rho v_\eta}{v}%\sqrt{\frac{v_\rho^2}{v_\eta^2}+1}} 
		& \frac{v_\rho^2-v_\eta^2}{v}\\%_\rho \sqrt{\frac{v_\eta^2}{v_\rho^2}+1}} \\
		%\frac{2 v_\rho}{\sqrt{\frac{v_\rho^2}{v_\eta^2}+1}} 
		\frac{2 v_\rho v_\eta}{v}& 0 & 0 \\
		%\frac{v_\rho^2-v_\eta^2}{v_\rho \sqrt{\frac{v_\eta^2}{v_\rho^2}+1}} 
		\frac{v_\rho^2-v_\eta^2}{v} & 0 & -\frac{v_\xi \left(v_\eta^2+v_\rho^2\right)}{v_\eta v_\rho} \\
	\end{array}
	\right)\,. \label{MCPeven2C}
\ee
If $v_\eta \approxeq v_\rho$, the third element of the first row (as well as the first element of the third row) in the matrix $M_{e2s_{(1)}}$ can be ignored. The matrix $M_{e2s_{(1)}}$ is then rewritten as:
\be
M_{e2r_{(1)}} = -\frac{B v_\chi}{2}\left(
\begin{array}{ccc}
	-\frac{v_\eta v_\rho}{v_\xi} & \frac{2 v_\rho v_\eta}{v}%\sqrt{\frac{v_\rho^2}{v_\eta^2}+1}} 
	& %\frac{v_\rho^2-v_\eta^2}{v_\rho \sqrt{\frac{v_\eta^2}{v_\rho^2}+1}} 
	0 \\
%	\frac{2 v_\rho}{\sqrt{\frac{v_\rho^2}{v_\eta^2}+1}}
\frac{2 v_\rho v_\eta}{v} & 0 & 0 \\
	%\frac{v_\rho^2-v_\eta^2}{v_\rho \sqrt{\frac{v_\eta^2}{v_\rho^2}+1}} 
	0& 0 & -\frac{v_\xi v^2%\left(v_\eta^2+v_\rho^2\right)
	}{v_\eta v_\rho} \\
\end{array}
\right)\,.
\ee
The $2 \times 2$ block on the left top of $M_{e2r_{(1)}}$ is rewritten as:
\be
M_{e2r_{(1)}} = -\frac{B v_\chi}{2}\left(
\begin{array}{ccc}
	-\frac{v_\eta v_\rho}{v_\xi} & \frac{2v_\rho v_\eta}{v}\\%\sqrt{\frac{v_\rho^2}{v_\eta^2}+1}} 	 \\
	\frac{2 v_\rho v_\eta}{v}%\sqrt{\frac{v_\rho^2}{v_\eta^2}+1}} 
	& 0  \\
\end{array}
\right)\,.
\ee
This $M_{e2r_{(1)}}$ matrix is diagonalized by the matrix:
\be
U_{\beta_{3_{(1)}}} = \left( 
\begin{array}{cc}
	\cos \beta_{3_{(1)}} & \sin \beta_{3_{(1)}} \\ 
	-\sin \beta_{3_{(1)}} & \cos \beta_{3_{(1)}} \\  
\end{array}
\right)\,.
\ee
Then, we receive a $3 \times 3$ rotational matrix with the form below:
\be
U_{\beta_{3s_{(1)}}}= \left(\begin{array}{ccc}
	 \cos \beta_{3_{(1)}} &\sin \beta_{3_{(1)}} &0 \\
	 \sin \beta_{3_{(1)}}  & \cos \beta_{3_{(1)}} &0\\ 
	  0 &0 &1 \\
	\end{array}
\right)\,.
\ee
with the mixing angle $\beta_{3_{(1)}}$ defined as:
\be
 \tan \beta_{3_{(1)}} = \frac{\sqrt{v^2+8v_\xi^2-v\sqrt{v^2+16v_\xi^2}}}{2\sqrt{2}v_\xi}\,.
\ee
Then, the physical states are:
\bea 
\mathcal{H}_{3_{(1)}} &=& R_\xi \cos \beta_{3_{(1)}} + R_\rho \sin \beta_{3_{(1)}}\,,\nn\\
\mathcal{H}_{4_{(1)}} &=& -R_\xi \sin \beta_{3_{(1)}} + R_\rho \cos \beta_{3_{(1)}}\,.
\eea
The masses of these two physical states are:
\be
m^2_{\mathcal{H}_{{3,4}_{(1)}}}= -\frac{B v_\chi v_\rho v_\eta}{4v v_\xi}\left( \sqrt{v^2+16v_\xi^2} \pm v\right)\,.
\ee
Since $B \sim \frac{1}{v_\chi}$, $\mathcal{H}_{4_{(1)}}$ acquires a mass at the same scale as SMLHB. The mass of $\mathcal{H}_{3_{(1)}}$ could be in the TeV or subTeV scale. %These masses are reasonale in comparision with the condition of the two Higgs doublets model \cite{2higgs}.
Finally, the matrix that diagonalizes the full $3\times 3$ matrix $	M^2_{even2b_{(1)}}$ is:
\bea
U_{even_{2b_{(1)}}}&=&U_{\beta_{3s_{(1)}}}.U_{\beta_{2s_{(1)}}}\nn\\
&=& \left(
\begin{array}{ccc}
	\cos \beta_{3_{(1)}} & \sin \beta_{2_{(1)}}
	\sin \beta_{3_{(1)}} & -\cos \beta_{2_{(1)}} \sin \beta_{3_{(1)}} \\
	-\sin \beta_{3_{(1)}} & \sin \beta_{2_{(1)}} \cos \beta_{3_{(1)}} & -\cos
	\beta_{2_{(1)}} \cos \beta_{3_{(1)}} \\
	0 & \cos \beta_{2_{(1)}} & \sin \beta_{2_{(1)}} \\
\end{array}
\right)\,.
\eea
\section{Diagonalizing the mass mixing matrix of the CP-even sectors in the model 2}\label{Diag2}

In the basis $(R_\phi, R_\rho, R_\eta^1)$, the mass squared matrix has the form: 
\begin{eqnarray}
M^{2}_{even_{2b_{(2)}}}= \left(
\begin{array}{ccc}
2\lambda _{35}
v_\phi^2 & \lambda _{38} v_\rho v_{\phi} & \lambda _{37} v_\eta v_\phi \\

\lambda _{38} v_\rho v_\phi & \Binh{-}\frac{f v_\eta
	v_\chi}{\sqrt{2} v_\rho}+2 \lambda _3 v_{\rho}^2 & \lambda _{10} v_\eta v_\rho-\frac{f
	v_\chi}{\sqrt{2}} \\

\lambda _{37} v_\eta v_\phi & \lambda _{10}
v_\eta v_\rho-\frac{f v_\chi}{\sqrt{2}} &
-\frac{f v_\rho v_\chi}{\sqrt{2} v_\eta}+2
\lambda _2 v_\eta^2 \\
\end{array}
\right)\,.  \label{2MCPeven1b}
\end{eqnarray}
 The $2 \times 2$ block on the right bottom of the matrix $M^{2}_{even_{2b_{(2)}}}$  in Eq. \eqref{2MCPeven1b} can be be rewritten as:
	\be
M^{2}_{even_{2bs_{(2)}}} =	\left(
	\begin{array}{cc}
		2 \lambda _3 v_\rho^2-\frac{f v_\eta v_\chi}{\sqrt{2}
			v_\rho} & \frac{f v_\chi}{\sqrt{2}}+\lambda _{10}
		v_\eta v_\rho \\
		\frac{f v_\chi}{\sqrt{2}}+\lambda _{10} v_\eta v_\rho
		& 2 \lambda _2 v_\eta^2-\frac{f v_\rho v_\chi}{\sqrt{2} v_\eta} \\
	\end{array}
	\right)=-\frac{f v_\chi}{\sqrt2} \left( 
	\begin{array}{cc}
		\frac{v_\eta }{ v_\rho} & 1 \\ 
		1 & \frac{v_\rho }{ v_\eta}
	\end{array}
	\right)\,.
	\ee
	Then we get a massive state $h_{(2)} = \cos \beta_{2_{(2)}} R_\rho +\sin \beta_{2_{(2)}} R_\eta^1 $. This might be the SM-like Higgs boson (SMLHB) whose mass is:
	\be
	m_{h_{(2)}}^2 %= \frac{-f v_\chi v_\xi}{2}\left(\frac{v_\rho}{v_\eta} + \frac{v_\eta}{v_\rho}\right) 
	= \frac{-f v_\chi  v^2}{\sqrt2 v_\rho v_\eta}\,,
	\ee
	where $v^2 = v_\rho^2 + v_\eta^2 = (246$ GeV$)^2$ and the parameter $f$ should be negative with its absolute value is $\sim \frac{1}{v_\chi}$. The mixing angle $\beta_{2_{(2)}}$ is defined by the equation $\tan \beta_{2_{(2)}} = \tan \beta_{2_{(1)}} =-\cot \al_{1_{(1)}}$.

Correspondingly, we get a $3 \times 3$ matrix as below:
\be
U_{\beta_{2_{(2)}}}= \left(\begin{array}{ccc}
	1 & 0 &0 \\
	0 & \cos \beta_{2_{(2)}} &\sin \beta_{2_{(2)}} \\
	0 & \sin \beta_{2_{(2)}}  & \cos \beta_{2_{(2)}} 
\end{array}
\right)\,.
\ee
The matrix $U_{\beta_{2_{(2)}}}$ transforms the matrix $M^{2}_{even_{2b_{(2)}}}$ into another form:
\be
M^{2}_{even_{2br_{(2)}}}= \left(
\begin{array}{ccc}
	2 \lambda _{35} v_\phi^2 & \frac{v_\phi \left(\lambda _{37}
		v_\eta^2+\lambda _{38} v_\rho^2\right)}{v_\eta
		\sqrt{\frac{v_\rho^2}{v_\eta^2}+1}} & \frac{\left(\lambda
		_{37}-\lambda _{38}\right) v_\eta v_\phi}{\sqrt{\frac{v_\eta^2}{v_\rho^2}+1}} \\
	\frac{v_\phi \left(\lambda _{37} v_\eta^2+\lambda _{38}
		v_\rho^2\right)}{v_\eta \sqrt{\frac{v_\rho^2}{v_\eta^2}+1}} & \frac{2 \left(\lambda _2 v_\eta^4+\lambda _{10} v_\eta^2 v_\rho^2+\lambda _3 v_\rho^4\right)}{v_\eta^2+v_\rho^2} & \frac{2 \lambda _2
		v_\eta^2+\lambda _{10} (v_\rho-v_\eta)
		(v_\eta+v_\rho)-2 \lambda _3 v_\rho^2}{\sqrt{\frac{v_\eta^2}{v_\rho^2}+1}
		\sqrt{\frac{v_\rho^2}{v_\eta^2}+1}} \\
	\frac{\left(\lambda _{37}-\lambda _{38}\right) v_\eta v_\phi}{\sqrt{\frac{v_\eta^2}{v_\rho^2}+1}} & \frac{2 \lambda _2
		v_\eta^2+\lambda _{10} (v_\rho-v_\eta)
		(v_\eta+v_\rho)-2 \lambda _3 v_\rho^2}{\sqrt{\frac{v_\eta^2}{v_\rho^2}+1}
		\sqrt{\frac{v_\rho^2}{v_\eta^2}+1}} & \frac{2 \left(\lambda
		_2+\lambda _3-\lambda _{10}\right) v_\eta^2 v_\rho^2}{v_\eta^2+v_\rho^2}-\frac{f v_\chi}{
		\left(v_\eta^2+v_\rho^2\right)}{\sqrt{2} v_\eta
		v_\rho} \\
\end{array}
\right).
\ee
With the assumption that $\lambda _{37} \approx \lambda _{38}$, $\lambda _{2}\approx\lambda _{3} \approx \lambda _{10}$ and $v_\rho \approxeq v_\eta$, the matrix $M^{2}_{even_{2br_{(2)}}}$ might be rewritten as below:
\be
M^{2}_{even_{2bt_{(2)}}}=\left(
\begin{array}{ccc}
	2 \lambda _{35} v_\phi^2 & \lambda _{37} v_\rho  v_\phi & 0 \\
	 \lambda _{37} v_\rho  v_\phi & \frac{3}{2}
	f^2 \lambda _2 v_\rho^2 v_\chi^2 & 0 \\
	0 & 0 &  \lambda _2 v_\rho^2-\sqrt{2} f v_\chi \\
\end{array}
\right).
\ee
The $2 \times 2$ block on the left top of the matrix $M^{2}_{even_{2bt_{(2)}}}$ gives two massive bosons whose physical states are related to the gauge eigenstates as:
\be
%U_{\beta_{3{(2)}}} 
\left( 
\begin{array}{c} \mathcal{H}_{3_{(2)}} \\ \mathcal{H}_{4_{(2)}}
\end{array}
\right)= \left( 
\begin{array}{cc}
	\cos \beta_{3_{(2)}} & \sin \beta_{3_{(2)}} \\ 
	-\sin \beta_{3_{(2)}} & \cos \beta_{3_{(2)}} \\ 
\end{array}
\right)\left( 
\begin{array}{c} R_\phi \\ R_\rho
\end{array}
\right)\,,
\ee
where the mixing angle $\beta_{3_{(2)}}$ is defined by:
\be
\tan \beta_{3_{(2)}} = \frac{2 \la_{35} v_\phi^2 - 3 \la_2 v_\rho^2-\sqrt{\left(3 \la_2 v_\rho^2 - 2 \la_{35} v_\phi^2\right)^2+8 \la_{37}^2 v_\rho^2 v_\phi^2}}{2\sqrt2 \la_{37} v_\rho v_\phi}\,.
\ee
The masses of these two scalars are:
\be
m^2_{\mathcal{H}_{3,4_{(2)}}} = \frac{3 \lambda _2 v_\rho^2}{2}+\lambda _{35} v_\phi^2\pm\frac{1}{2} \sqrt{\left(3 \lambda _2
	v_\rho^2-2 \lambda _{35} v_\phi^2\right){}^2+8 \lambda
	_{37}^2 v_\rho^2 v_\phi^2}\,.
\ee
Finally, the matrix which diagonalizes the matrix $	M^2_{even2b_{(1)}}$ is:
\bea
U_{even_{2b_{(2)}}}&=&U_{\beta_{3s_{(1)}}}.U_{\beta_{2s_{(1)}}}\nn\\
&=& \left(
\begin{array}{ccc}
	\cos \beta_{3_{(2)}} & \sin \beta_{2_{(2)}}
	\sin \beta_{3_{(2)}} & -\cos \beta_{2_{(2)}} \sin \beta_{3_{(2)}} \\
	-\sin \beta_{3_{(2)}} & \sin \beta_{2_{(2)}} \cos \beta_{3_{(2)}} & -\cos
	\beta_{2_{(2)}} \cos \beta_{3_{(2)}} \\
	0 & \cos \beta_{2_{(2)}} & \sin \beta_{2_{(2)}} \\
\end{array}
\right)\,.
\eea

\bibliographystyle{utphys}
\bibliography{331IS2024}
\end{document}